\pdfoutput=1  
\documentclass[acmsmall]{acmart}

\usepackage[linesnumbered,ruled,vlined]{algorithm2e}
\usepackage{tabularx}
\usepackage{blindtext}
\usepackage{multirow}
\usepackage{stfloats}
\usepackage{enumitem}

\usepackage{textcomp}
\usepackage{color}
\usepackage{xcolor}
\usepackage{xspace}
\usepackage{balance}

\usepackage{soul}

\usepackage{pgfplots}
\usepackage{pgfplotstable}
\pgfplotsset{compat=1.17}

\usepackage{tikz}
\usetikzlibrary{patterns}

\usepackage{subcaption}
\usepackage{graphicx}
\usepackage{wasysym}

\setlist[itemize]{leftmargin=1.5em, labelsep=0.5em}
\AtBeginDocument{%
  }

\setcopyright{acmlicensed}
\acmJournal{PACMMOD}
\acmYear{2025} \acmVolume{3} \acmNumber{3 (SIGMOD)} \acmArticle{208} \acmMonth{6}\acmDOI{10.1145/3725345}

\begin{document}

\newcommand{\stitle}[1]{\noindent{\textbf{#1}}}
\newcommand{\kw}[1]{{\ensuremath {\mathsf{#1}}}\xspace}
\newcommand{\kwnospace}[1]{{\ensuremath {\mathsf{#1}}}}
\newcommand{\todo}[1]{\textcolor{red}{$\Rightarrow:$ #1}}
\newcommand{\revise}[1]{#1}
\newcommand{\delete}[1]{}
\newcommand{\notation}[1]{}

\newcommand*\circled[1]{\tikz[baseline=(char.base)]{
            \node[shape=circle,draw,inner sep=1pt] (char) {#1};}}

\newcommand{\update}[1]{\textcolor{red}{$\Rightarrow:$ #1}}

\definecolor{c1}{RGB}{42,99,172} %
\definecolor{c2}{RGB}{255,88,93}
\definecolor{c3}{RGB}{255,181,73}
\definecolor{c4}{RGB}{119,71,64} %
\definecolor{c5}{RGB}{228,123,121} %
\definecolor{c6}{RGB}{208,167,39} %
\definecolor{c7}{RGB}{0,51,153}
\definecolor{c8}{RGB}{56,140,139} 
\definecolor{c9}{RGB}{0,0,0} 

\newcommand{\sssp}{\kw{SSSP}}
\newcommand{\pr}{\kw{PR}}
\newcommand{\tc}{\kw{TC}}
\newcommand{\bc}{\kw{BC}}
\newcommand{\bfs}{\kw{BFS}}
\newcommand{\clique}{\kw{KC}}
\newcommand{\core}{\kw{CD}}
\newcommand{\cc}{\kw{WCC}}
\newcommand{\lcc}{\kw{LCC}}
\newcommand{\wcc}{\kw{WCC}}
\newcommand{\lpa}{\kw{LPA}}
\newcommand{\cd}{\kw{CD}}
\newcommand{\kc}{\kw{KC}}
\newcommand{\khop}{\kw{K}-\kw{Hop}}
\newcommand{\cluster}{\kw{Cluster}}

\newcommand{\vc}{\kw{vertex}-\kw{centric}}
\newcommand{\vcc}{\kw{Vertex}-\kw{centric}}
\newcommand{\ec}{\kw{edge}-\kw{centric}}
\newcommand{\blc}{\kw{block}-\kw{centric}}
\newcommand{\blcc}{\kw{Block}-\kw{centric}}
\newcommand{\sgc}{\kw{subgraph}-\kw{centric}}

\newcommand{\graphx}{\kw{GraphX}}
\newcommand{\power}{\kw{PowerGraph}}
\newcommand{\flash}{\kw{Flash}}
\newcommand{\grape}{\kw{Grape}}
\newcommand{\pregel}{\kw{Pregel+}}
\newcommand{\ligra}{\kw{Ligra}}
\newcommand{\gthinker}{\kw{G}-\kw{thinker}}
\newcommand{\pregell}{\kw{Pregel}}
\newcommand{\scattergather}{\kw{Scatter}-\kw{Gather}}
\newcommand{\scattercombine}{\kw{Scatter}-\kw{Combine}}
\newcommand{\xstream}{\kw{X}-\kw{Stream}}
\newcommand{\graphchi}{\kw{Graphchi}}
\newcommand{\chaos}{\kw{Chaos}}
\newcommand{\blogel}{\kw{Blogel}}
\newcommand{\arabesque}{\kw{Arabesque}}
\newcommand{\fractal}{\kw{Fractal}}
\newcommand{\autoMine}{\kw{AutoMine}}
\newcommand{\peregrine}{\kw{Peregrine}}
\newcommand{\vrdd}{\kw{VertexRDD}}

\newcommand{\ldbcdg}{\kwnospace{LDBC}-\kw{DG}}
\newcommand{\stgt}{\kw{S3G2}}
\newcommand{\ldbc}{\kw{LDBC~Graphalytics}}
\newcommand{\graphfive}{\kw{Graph500}}
\newcommand{\fftdg}{\kw{FFT}-\kw{DG}}
\title{Revisiting Graph Analytics Benchmark}

\author{Lingkai Meng}
\authornote{These authors contribute equally to this work and are listed in alphabetical order by last name.}
\affiliation{%
  \institution{Shanghai Jiao Tong University}
  \city{Shanghai}
  \country{China}}
\email{mlk123@sjtu.edu.cn}

\author{Yu Shao}
\authornotemark[1]
\affiliation{%
  \institution{East China Normal University}
  \city{Shanghai}
  \country{China}}
\email{yushao@stu.ecnu.edu.cn}

\author{Long Yuan}
\authornote{Corresponding author.}
\affiliation{%
  \institution{Wuhan University of Technology}
  \city{Wuhan}
  \country{China}}
\email{longyuanwhut@gmail.com}

\author{Longbin Lai}
\affiliation{%
  \institution{Alibaba Group}
  \city{Hangzhou}
  \country{China}}
\email{longbin.lailb@alibaba-inc.com}

\author{Peng Cheng}
\affiliation{%
  \institution{Tongji University}
  \city{Shanghai}
  \country{China}}
\email{cspcheng@tongji.edu.cn}

\author{Xue Li}
\affiliation{%
  \institution{Alibaba Group}
  \city{Hangzhou}
  \country{China}}
\email{youli.lx@alibaba-inc.com}

\author{Wenyuan Yu}
\affiliation{%
  \institution{Alibaba Group}
  \city{Hangzhou}
  \country{China}}
\email{wenyuan.ywy@alibaba-inc.com}

\author{Wenjie Zhang}
\affiliation{%
  \institution{University of New South Wales}
  \city{Sydney}
  \country{Australia}}
\email{wenjie.zhang@unsw.edu.au}

\author{Xuemin Lin}
\affiliation{%
  \institution{Shanghai Jiao Tong University}
  \city{Shanghai}
  \country{China}}
\email{xuemin.lin@gmail.com}

\author{Jingren Zhou}
\affiliation{%
  \institution{Alibaba Group}
  \city{Hangzhou}
  \country{China}}
\email{jingren.zhou@alibaba-inc.com}

\renewcommand{\shortauthors}{Lingkai Meng et al.}

\begin{abstract}

The rise of graph analytics platforms has led to the development of various benchmarks for evaluating and comparing platform performance. However, existing benchmarks often fall short of fully assessing performance due to limitations in core algorithm selection, data generation processes (and the corresponding synthetic datasets), as well as the neglect of API usability evaluation. To address these shortcomings, we propose a novel graph analytics benchmark. First, we select eight core algorithms by extensively reviewing both academic and industrial settings. Second, we design an efficient and flexible data generator and produce eight new synthetic datasets as the default datasets for our benchmark. Lastly, we introduce a multi-level large language model (LLM)-based framework for API usability evaluation-the first of its kind in graph analytics benchmarks. We conduct comprehensive experimental evaluations on existing platforms ({\graphx, \power, \flash, \grape, \pregel, \ligra, and }\gthinker). The experimental results demonstrate the superiority of our proposed benchmark.

\end{abstract}

\ccsdesc[500]{Information systems~Database performance evaluation}

\keywords{Graph Analytics Benchmarks; Data Generator; API Usability Evaluation}

\received{October 2024}
\received[revised]{January 2025}
\received[accepted]{February 2025}

\maketitle

\section{Introduction}
\label{sec:intro}

Graph analytics platforms, such as \graphx~\cite{gonzalez2014graphx}, \power~\cite{gonzalez2012powergraph}, \flash~\cite{li2023flash}, \grape~\cite{fan2018parallelizing}, \pregel~\cite{yan2015effective,yan2014pregel,lu2014large}, \ligra~\cite{shun2013ligra}, and \gthinker \cite{yan2020g}, have been developed to facilitate large-scale graph data analysis. These platforms vary significantly in their design principles, architectures, and implementation details, which raises an important and practical question: how to evaluate graph analytics platforms and provide guidance on selecting the most appropriate one for specific environments.

Accordingly, graph analytics benchmarks like \ldbc \cite{DBLP:journals/pvldb/IosupHNHPMCCSAT16} and \graphfive \cite{graph500} are commonly used to evaluate platform performance. These benchmarks provide a standardized framework for assessment, typically consisting of two main components: core algorithms and datasets. Core algorithms measure the efficiency of various graph processing tasks, offering a consistent basis for comparing platforms. Datasets aim to represent diverse graph characteristics found in real-world applications. However, due to high collection costs and privacy concerns, real-world datasets are often limited. As a result, benchmarks frequently use data generators to produce synthetic datasets that mimic real-world graphs.

\stitle{Motivation.} Considering the goals and key components of graph analytics benchmarks, an ideal benchmark should select the minimal number of algorithms while ensuring they are representative and diverse enough to effectively distinguish the performance of various graph analytics platforms. The data generator should also efficiently produce synthetic datasets, allowing users to control input parameters to generate graphs with diverse characteristics. Unfortunately, existing benchmarks fall short of these ideals. Take the most popular \ldbc benchmark \cite{DBLP:journals/pvldb/IosupHNHPMCCSAT16} as an example. Regarding core algorithms, \ldbc selects PageRank (\pr)~\cite{page1999pagerank}, Breadth First Search (\bfs)~\cite{lee1961algorithm}, Single Source Shortest Path (\sssp)~\cite{DBLP:journals/nm/Dijkstra59}, Weakly Connected Component (\wcc)~\cite{tarjan1984worst}, Label Propagation Algorithm (\lpa)~\cite{raghavan2007near}, Local Clustering Coefficient (\lcc)~\cite{holland1971transitivity} as its core algorithms. However, as analyzed in Section \ref{sec:selected_algorithms}, these six algorithms lack diversity, with most having linear time complexity, limiting their ability to reveal platform performance bottlenecks. Regarding dataset generation, \ldbc's data generator employs a sampling-based strategy to generate datasets, which, as examined in Section \ref{sec:generator}, suffers from efficiency issues due to numerous failed trials. Additionally, the generator only allows control over the scale of datasets, lacking flexibility in mimicking real-world datasets of different characteristics.

Moreover, current benchmarks predominantly assess and compare performance metrics such as execution time, throughput, and scalability. However, these benchmarks often overlook the usability of the application programming interfaces (APIs) provided by graph analytics platforms. APIs play a crucial role in enhancing developer productivity, minimizing errors, and fostering wider platform adoption. The success of a graph analytics platform is significantly tied to the usability of its APIs, as highlighted in studies~\cite{DBLP:journals/csr/RaufTP19,DBLP:conf/icse/Myers17}. Consequently, API usability should be a key component in the design of graph analytics benchmarks. Yet, existing benchmarks typically neglect API usability, relying instead on simplistic measures like the number of lines of code to implement tasks \cite{6976096,DBLP:conf/kbse/NamHMMV19,watson2009improving}. Such metrics are insufficient for a comprehensive evaluation of API usability.

Motivated by these gaps, in this paper, we revisit the problem of graph analytics benchmarks and aim to develop a new benchmark that includes a diverse set of core algorithms, an efficient and flexible data generator, and a rigorous method for evaluating API usability.

\stitle{Our Approach.} We address the three major limitations in existing graph analytics benchmarks. For the core algorithms, we conduct an extensive review of both academic research and industrial applications, selecting eight algorithms (\pr, \sssp, \lpa, \wcc, Triangle Counting (\tc), Betweenness Centrality (\bc), $k$-Clique (\kc), Core Decomposition (\cd)) based on their popularity, time complexity, and diversity of algorithmic topics. For the data generator, we redesign the edge sampling strategy and introduce a failure-free generator that eliminates the inefficiencies caused by failed trials in \ldbc. Additionally, recognizing the critical impact of graph density and diameter on platform performance \cite{yan2015effective,fan2018parallelizing,li2023flash}, we develop a mechanism that allows users to  control these two attributes of the generated datasets, significantly enhancing the flexibility of the data generator.  For API usability evaluation, we harness the capabilities of large language models (\kwnospace{LLM}s) and propose a multi-level \kw{LLM}-based evaluation framework that simulates programmers of varying expertise levels, providing a comprehensive assessment of API usability. Table~\ref{tab:benchmark-comparison} compares our proposed benchmark with existing graph analytics benchmarks regarding core algorithms, controllable attributes of the synthetic datasets, and the evaluation metrics.

\begin{table}[t]
\centering
\caption{Comparison of benchmarks for graph analytics platforms}
\label{tab:benchmark-comparison}
\vspace{-0.5em}
\small
\resizebox{\textwidth}{!}{
\begin{tabular}{>{\centering\arraybackslash}p{3cm}|>{\centering\arraybackslash}p{2.8cm}|>{\centering\arraybackslash}p{2.8cm}|>{\centering\arraybackslash}p{3cm}|>{\centering\arraybackslash}p{1.8cm}}
\hline 
\multirow{2}{*}{\textbf{Benchmarks}} & \multirow{2}{*}{\textbf{Core Algorithms}} & \multirow{2}{*}{\textbf{Synthetic Datasets}} & \multicolumn{2}{c}{\textbf{Evaluation Metrics}} \\ \cline{4-5}
 & & & \textbf{Performance} & \textbf{Usability} \\ \hline
\kw{Graph500} \cite{graph500}          & \bfs, \sssp                             & Scale                & Timing, Throughput               & --- \\ \cline{1-5}
\multirow{2}{*}{\kw{WGB}~\cite{DBLP:conf/wbdb/AmmarO13}}       & \khop~\cite{wang2009mobility}, \sssp, \pr, \wcc, \cluster~\cite{schaeffer2007graph}                           & \multirow{2}{*}{Scale, Density}          & \multirow{2}{*}{Timing, Scalability}              & \multirow{2}{*}{---} \\ \cline{1-5}
\kw{BigDataBench} \cite{DBLP:conf/hpca/WangZLZYHGJSZZLZLQ14}       & \bfs, \pr, \wcc, \cluster                    & Scale            &  Timing, Throughput, MIPS, cache MPKI              & --- \\ \cline{1-5}
\ldbc \cite{DBLP:journals/pvldb/IosupHNHPMCCSAT16}  & \pr, \bfs, \sssp, \wcc, \lpa, \lcc               & Scale                & Timing, Throughput, Scalability, Robustness & --- \\ \cline{1-5}
\kw{{Ours}}   & \pr, \sssp, \tc, \bc, \kc, \cd, \lpa, \cc & Scale, Density, Diameter &  Timing, Throughput, Scalability, Robustness & LLM-based \\
\hline
\end{tabular}
}
\vspace{-1em}
\end{table}

\stitle{Contributions.} In this paper, we make the following contributions:

\begin{itemize}[leftmargin=*]

\item \notation{A.(4)} \revise{We revisit existing graph analytics benchmarks, identify their limitations, and introduce a more comprehensive benchmark that supports a broader range of algorithms and datasets, along with a richer set of evaluations, including the assessment of API usability.}


\item \notation{A.(4)} \revise{We select eight representative core algorithms, spanning a wide range of algorithm popularity, algorithm diversity and computing models, to ensure a comprehensive and realistic evaluation of platform performance (Section~\ref{sec:selected_algorithms}).}

\item We design an efficient and flexible data generator and produce eight synthetic datasets using our new data generator that serve as the default for our benchmark (Section \ref{sec:generator}).

\item We propose an \kw{LLM}-based evaluation framework to assess the API usability. To the best of our knowledge, this is the first graph analytics benchmark that includes a rigorous method for evaluating API usability  (Section \ref{sec:evaluator}).
\item We conduct a comprehensive experimental evaluation on existing graph analytics platforms using the newly proposed benchmark, and the experimental results demonstrate the superiority of our benchmark (Section \ref{sec:exp_result}).
\end{itemize}
\section{Related Work}
\label{sec:related_work}



\stitle{Benchmarks for  Graph Analytics Platforms.} Benchmarking graph analytics platforms has attracted lots of attention in recent years, and many general-purpose benchmarks have been proposed in the literature. \kw{Graph500} \cite{graph500} uses \kw{BFS} and \kw{SSSP} algorithms on synthetic datasets generated by the \kw{Kronecker} graph generator to measure platform performance in terms of timing and throughput. \kw{WGB} \cite{DBLP:conf/wbdb/AmmarO13} provides an efficient data generator for creating dynamic graphs similar to real-world graphs, offering a universal benchmark for graph analytics platforms. \kw{BigDataBench} \cite{DBLP:conf/hpca/WangZLZYHGJSZZLZLQ14} aims to provide a comprehensive benchmarking suite for data systems, encompassing various data types, including graph data, text, and tables, and focusing on performance, energy efficiency, and cost-effectiveness under diverse workloads and data scenarios. \kw{LDBC} \cite{DBLP:journals/pvldb/IosupHNHPMCCSAT16} is the most popular benchmark for graph analysis systems, offering datasets and evaluation metrics for various tasks, establishing itself as the state-of-the-art tool for assessing graph analytics platforms.

\delete{
Besides general-purpose benchmarks, special-purpose graph analytics benchmarks are also studied in the literature. 
\kw{XGDBench} \cite{DBLP:conf/cloudcom/DayarathnaS12} benchmarks graph databases in cloud computing systems. 
\kw{LinkBench} \cite{DBLP:conf/sigmod/ArmstrongPBC13} emulates social graph database workloads for realistic database performance benchmarks. 
\kw{GraphBIG} \cite{DBLP:conf/sc/NaiXTKL15} includes metrics for GPUs, assessing different hardware architectures' handling of graph workloads. 
\kw{MiniVite} \cite{DBLP:conf/sc/GhoshHTKG18} focuses on irregular complex computations, using the Louvain algorithm to evaluate graph analytics platforms. 
\kw{GraphTides} \cite{DBLP:conf/grades/ErbMKSCMP18} benchmarks stream-based graph analytics platforms. 
The Open Graph Benchmark (\kw{OGB}) \cite{DBLP:conf/nips/HuFZDRLCL20} focuses on machine learning tasks involving graph data.
}


\delete{
\stitle{Performance Study for  Graph Analytics Platforms.} Performance studies for graph analytics platforms are prevalent in the literature. Salim Jouili et al. \cite{DBLP:conf/socialcom/JouiliV13} empirically compare graph databases Neo4j, OrientDB, Titan, and DEX, analyzing performance on common graph operations like neighborhood exploration, shortest path finding, and vertex retrieval with specific properties. Minyang Han et al. \cite{DBLP:journals/pvldb/HanDAOWJ14} compare Pregel-like graph processing systems such as Apache Giraph, GPS, Mizan, and GraphLab. Yi Lu et al. \cite{DBLP:journals/pvldb/LuCYW14} broaden benchmarks to include algorithms like PageRank, HashMin, and more, using diverse datasets (social, web, RDF, road networks) to assess timing and scalability. Other studies \cite{DBLP:journals/pvldb/AmmarO18,DBLP:journals/jcsm/0014YXZS014} conduct extensive evaluations of distributed graph analytics platforms, examining performance and scalability on large graphs.
}


\stitle{Synthetic Graph Data Generators in Benchmarks.} 
The graph data generator creates synthetic graph data that simulates real-world network structures and behaviors for use in benchmarking and analysis.
\notation{A.(2)} \revise{Traditional \textsf{Erdős}-\textsf{Rényi} generator~\cite{erdos1960evolution} iteratively selects two random vertices to generate an edge. However, the generated graph has a low clustering coefficient and follows a Poisson degree distribution, which is contrary to real-world networks. Two improved generators, \kw{Watts}-\kw{Strogatz}~\cite{watts1998collective} and \textsf{Barabási}-\kw{Albert}~\cite{barabasi1999emergence}, can generate highly-clustered and power-law graphs, respectively, while the \kw{Kronecker}~\cite{leskovec2010kronecker} generator used in \kw{Graph500} ~\cite{murphy2010introducing} can simultaneously follow these two properties.
The \kw{LFR3}~\cite{lancichinetti2008benchmark} generator allows for the tuning of parameters to control clusters and power-law distribution. The \ldbcdg generator can generate graphs with arbitrary degree distributions and better community similarity~\cite{erling2015ldbc}.
In addition to the general graph generator, the \kw{WGB} benchmark~\cite{DBLP:conf/wbdb/AmmarO13} adds a dynamic graph generator for evaluating systems under dynamic conditions.
Many generators are integrated in graph libraries such as \kw{NetworkX}~\cite{hagberg2008exploring} and \kw{Neo4j}-\kw{APOC}~\cite{neo4j-apoc}.
However, these generators are not efficient enough and cannot control the density and diameter of the generated graphs.}


\stitle{API Usability Evaluation.} Due to the importance of APIs in software development, many studies \cite{DBLP:journals/csr/RaufTP19,DBLP:conf/icse/Myers17,DBLP:conf/esem/PiccioniFM13,DBLP:conf/hcse/GrillPT12,DBLP:journals/infsof/Mosqueira-ReyAM18,DBLP:conf/chi/FarooqWZ10} have explored methods for evaluating API usability to guide API design. 
Existing works mainly rely on expert evaluations by project managers, developers, and testers, who review the API design and documentation to evaluate its usability and suggest improvements \cite{DBLP:journals/csr/RaufTP19,DBLP:conf/chi/FarooqWZ10}, 
while others gather insights through long-term tracking of developer feedback~\cite{mosqueira2018systematic}.  Brad A. Myers \cite{DBLP:conf/icse/Myers17} points out that better API usability requires designing APIs to effectively meet the needs of users at different levels, including novices, professionals, and end-user programmers. Therefore, some studies \cite{DBLP:conf/icse/Myers17,DBLP:conf/esem/PiccioniFM13} have adopted the method of collecting feedback from users with varying levels of expertise to evaluate API usability. Clearly, this approach is costly and difficult to scale for large benchmark testing.

\section{Core Algorithms}
\label{sec:selected_algorithms}

Algorithms play a vital role in benchmarking distributed graph processing platforms.
\kw{LDBC}~\cite{DBLP:journals/pvldb/IosupHNHPMCCSAT16}, the most
popular benchmark for graph processing platforms, selects six algorithms (\bfs, \pr, \wcc, \lpa, \lcc and \sssp) frequently mentioned in academic papers as its core algorithms.
\revise{However, \kw{LDBC}'s algorithm set lacks diversity, as most of the algorithms focus on community detection and traversal, while ignoring the suitability for different computing models.}

\revise{To overcome these limitations, we consider multiple criteria when selecting core algorithms to ensure a comprehensive and representative benchmark, including (1) popularity in academic research and real-world application, (2) diversity in algorithm topics, and (3) suitability for distributed graph computing models. Based on these criteria, we select eight algorithms, including:}

\delete{
Algorithms play a vital role in benchmarking distributed graph processing platforms. Different algorithms stress various aspects of system performance, such as computation efficiency, memory usage, and communication overhead.
\kw{LDBC}~\cite{DBLP:journals/pvldb/IosupHNHPMCCSAT16}, as the most
popular benchmark for graph analysis platforms, selects six algorithms (\bfs, \pr, \wcc, \lpa, \lcc and \sssp) as its core algorithms, but the selection criteria are relatively narrow, focusing primarily on algorithms frequently mentioned in academic papers. Moreover, the \kw{LDBC}'s core algorithm set lacks diversity, as most of the algorithms have linear time complexity, limiting the potential to reveal performance bottlenecks of the platform.
Moreover, these algorithms mainly cover community detection and traversal, failing to represent the full range of challenges in distributed graph processing~\cite{meng2024survey}.
To overcome these limitations, we considered multiple criteria when selecting core algorithms to ensure a comprehensive and representative benchmark. The first criterion is popularity. Algorithms that frequently appear in academic research and industrial applications often reflect common real-world needs, ensuring that our benchmark is not only theoretically meaningful but also practically relevant. Another important criterion is computation workload and complexity. By selecting algorithms with varying time complexities, we can comprehensively test the platform's performance under different computational loads, ensuring that we can identify the platform's bottlenecks and strengths.
Finally, we emphasized diversity in algorithm topics, selecting algorithms from a wide variety of areas. This ensures a comprehensive evaluation of the platform's performance when facing different distributed graph processing challenges. Based on the previous criteria, we selected eight representative algorithms, including:
}

\begin{itemize}[leftmargin=*]

\item \textbf{PageRank (\pr)} measures the importance of each vertex based on the number and quality of its edges.

\item \textbf{Label Propagation Algorithm (\lpa)} finds communities by exchanging labels between vertices, using a semi-supervised approach.

\item \textbf{Single Source Shortest Path (\sssp)} finds the shortest paths from a given source vertex to all others based on the smallest sum of edge weights.

\item \textbf{Weakly Connected Component (\wcc)} identifies subgraphs where any two vertices are connected, ignoring edge direction.

\item \textbf{Betweenness Centrality (\bc)} quantifies the degree to which a vertex lies on the shortest paths between pairs of other vertices.

\item \textbf{Core Decomposition (\cd)} determines the coreness value of each vertex, indicating if it is part of a $k$-Core subgraph, where it is connected to at least $k$ other vertices.

\item \textbf{Triangle Counting (\tc)} counts the number of triangles in a graph, often used to measure the local clustering coefficient.

\item \textbf{$k$-Clique (\kc)} identifies all complete subgraphs with $k$ vertices.

\end{itemize}

\begin{table}[t]\centering

\caption{Popularity of selected core algorithms}
\label{tab:selected_algo}
\vspace{-0.5em}
\small
\begin{tabular}{c|c|c|c|c}
\hline
\textbf{Algorithms} & \textbf{\#Papers} & \textbf{DBLP} & \textbf{Google Scholar} & \textbf{  WoS} \\ \hline

\pr         & 28    & 1012 & 25400          & 4554           \\ \hline
\lpa         & 39    & 771  & 130000         & 1195         \\ \hline
\sssp       & 33    & 584  & 17800          & 2252            \\ \hline
\cc         & 26     & 835   & 17800          & 726658       \\ \hline
\bc         & 20    & 304  & 43900          & 5634           \\ \hline
\cd         & 29    & 179  & 126000         & 19499           \\ \hline
\tc         & 27    & 252  & 20500          & 1784            \\ \hline
\kc         & 31    & 352  & 41800          & 395           \\ \hline

\end{tabular}
\end{table}




\vspace{-0.3em}
\subsection{Algorithm Popularity}

Table~\ref{tab:selected_algo} presents statistics on the number of papers related to the selected algorithms, appearing in representative conferences and journals~\cite{meng2024survey}. It also shows their frequency of appearance in research engines like DBLP, Google Scholar, and Web of Science (WoS) over the past ten years.
\notation{A.(1)} \revise{{Moreover, the selected algorithms are not only frequently studied in academic papers but also have a wide range of practical applications. Here are just a few examples:}
\stitle{(1) Social Network Analysis.} \pr and \bc are used to detect important individuals in social networks by considering the influence of neighbors~\cite{jin2012lbsnrank, yin2019signed, gao2022mebc} and shortest paths~\cite{kourtellis2013identifying}, respectively. \lpa and \kc are used to detect communities~\cite{azaouzi2017evidential, ouyang2020clique}, while \cd can further reveal the hierarchical structure~\cite{malliaros2020core}.
\stitle{(2) Computer Vision.} \wcc can find connected pixels for unsupervised segmentation~\cite{wang1998stochastic, trevor2013efficient}. \lpa can utilize given labels for semantic segmentation~\cite{vernaza2017learning, mustikovela2016can}.
\stitle{(3) Biological Analysis.} Similar to the social network analysis, \bc and \pr are also adaptive to measure the importance of proteins~\cite{wang2024efficient, ivan2011web}, while \tc can measure similarity metrics like Jaccard similarity~\cite{buriol2006counting}. In epidemiology, \cd is suitable to reveal and predict epidemic outbreaks.
\stitle{(4)~Road Network Routing.} In road networks, \sssp and \wcc are used to compute shortest paths~\cite{forster2018faster, nanongkai2014distributed} and check the connection~\cite{xie2007measuring, wiedemann2000automatic} between vertices, respectively. \bc is applied to identify high-traffic areas frequently traversed by shortest paths~\cite{li2010betweenness, kazerani2009can}.
\stitle{(5) Recommendation Systems.} \pr can detect high influence items for recommendation~\cite{ma2008bringing}. \tc and \kc can detect cliques and triadic patterns to uncover close relationships for recommendation~\cite{wu2016link, carullo2015triadic, caramia2006mining}. \kc and \lpa are also used for fraud detection~\cite{berry2004emergent, li2022adaptive}.}

\begin{table}[t]\centering
        \caption{Workload and topics}
        \label{tab:selected_algo_comp}
        \vspace{-0.5em}
        \small
        
        \begin{tabular}{c|c|c|c|c}
        \hline
        \textbf{Algorithms} & \textbf{Workload} & \textbf{Topic} & \textbf{LDBC} & \textbf{Ours} \\ \hline

        \pr         & $O(k\cdot m)$    & Centrality   & \checkmark &  \checkmark       \\ \hline
        \lpa         & $O(k\cdot m)$    & Community Detection    & \checkmark &    \checkmark          \\ \hline
        \sssp       & $O(m+n\cdot\log{n})$ & Traversal    & \checkmark &    \checkmark        \\ \hline
        \cc         & $O(m+n)$     & Community Detection   & \checkmark &  \checkmark       \\ \hline
        \bc         & $O(n^3)$    & Centrality          & & \checkmark      \\ \hline
        \core         & $O(m+n)$    & Cohesive Subgraph    & &  \checkmark           \\ \hline
        \tc         & $O(m^{1.5})$    & Pattern Matching      & &  \checkmark      \\ \hline
        \clique         & $O(k^2\cdot n^k)$    & Pattern Matching    & &   \checkmark  \\ \hline
        \bfs         & $O(m+n)$    & Traversal    & \checkmark &     \\ \hline
        \lcc         & $O(m^{1.5})$    & Community Detection   &  \checkmark &      \\ \hline

        \end{tabular}
        
    \end{table}

\subsection{Algorithm Diversity}

\delete{These statistical results reveal that the \lpa boasts the highest number of papers (39) and the most appearances on Google Scholar (130,000), indicative of its popularity and extensive research in both academic and industry settings.
Similarly, \sssp and \pr algorithms are well-studied, with significant representation across various databases.}
Table~\ref{tab:selected_algo_comp} further compares the computational workload and topics of our core algorithms and \kw{LDBC}. Our algorithms span a range of complexities, from linear (\pr, \core, \cc, \lpa) to logarithmic-linear (\sssp), polynomial (\tc, \bc), and even higher-order complexities (\clique), providing a balanced coverage of different workload levels. In contrast, more than half of the algorithms (\pr, \lpa, \wcc, \bfs) in \kw{LDBC} are linear.
Regarding the topics, our core algorithms cover areas including centrality, community detection, traversal, cohesive subgraph, and pattern matching, highlighting the diversity of our algorithms, while \kw{LDBC} only includes centrality, community detection, traversal, with half of its algorithms focused on community detection.
Overall, our core algorithms are widely used in academia and industry, addressing real-world needs while covering diverse workloads, enabling our benchmark to identify platform bottlenecks and strengths effectively.

\vspace{-0.3em}
\revise{\subsection{Algorithm Suitability for Computing Models}
\label{sec:algorithm_suitability}

\notation{A.(3)} The performance of distributed graph algorithms is significantly impacted by the underlying computing model. In the following, we explore common graph computing models and showcase the varying suitability of the selected algorithms across these models.

\stitle{Graph Computing Models.} 
The Bulk Synchronous Parallelism (BSP) model, introduced by Valiant~\cite{valiant1990bridging}, is a popular computing model for parallel graph computation. It divides the computation into multiple supersteps, where each superstep consists of three phases: local computation, communication, and synchronization, for efficiently executing parallel tasks. 

On top of the BSP model, the \vc model was developed with the ``Think-Like-a-Vertex'' (TLAV) philosophy ~\cite{mccune2015thinking}. In \vc model, the vertex is the basic unit of computing and scheduling. Each vertex can process computation locally and send messages to other vertices. Users are only required to implement a vertex-based computation interface, while the platform will execute it over all vertices for a certain number of iterations or until it converges.
There are a few variants of \vc model such as Single-Phase (\pregell~\cite{malewicz2010pregel}, \pregel~\cite{yan2015effective}, \flash~\cite{10184838}, \ligra~\cite{DBLP:conf/ppopp/ShunB13}), \scattergather (Signal/Collect~\cite{stutz2010signal}), and \scattercombine (GRE~\cite{yan2014pregel}). Specially, Spark also supports \vc by providing Pregel-like interfaces, i.e., \vrdd in GraphX~\cite{gonzalez2014graphx}. \vcc models face potential challenges, including load imbalance, which becomes particularly problematic in power-law graphs, and significant communication overhead when deployed on large-scale clusters.

Some platforms build upon the \vc model by extending computation on edges or groups of vertices, leading to the development of \ec and \blc models, respectively. PowerGraph~\cite{gonzalez2012powergraph}, \xstream~\cite{roy2013x}, \graphchi~\cite{kyrola2012graphchi} and \chaos~\cite{roy2015chaos} provide an \ec model to execute tasks over edges to resolve load skew in power-law graphs and fully utilize sequential reading and writing of solid-state disk. Alternatively, \blogel~\cite{yan2014blogel} and \grape~\cite{fan2017grape} are \blc platforms that divide the graph into multiple blocks, such that vertex functions within the same block can exchange information without communication. However, \ec models do not support non-neighbor communication, \mbox{while it's often non-trivial to program with the \blc model.}

All the above computing models produce output sizes proportional to the graph size, making them unsuitable for graph-mining problems that may yield an exponential number of results~\cite{yan2017big, yan2024systems}. To address this, platforms like \arabesque~\cite{teixeira2015arabesque}, \fractal~\cite{dias2019fractal}, \autoMine~\cite{mawhirter2019automine}, \peregrine~\cite{jamshidi2020peregrine}, and \gthinker~\cite{yan2020g} adopt the \sgc model. In this model, the fundamental computing unit is the subgraph (e.g., a triangle) rather than vertices, vertex groups, or edges. Users need to define how to construct candidate subgraphs and implement a computation interface for each subgraph.


\stitle{Algorithm Categorizations.}
Different graph computing models can greatly influence algorithm efficiency,
as their diverse algorithmic requirements may align differently with the characteristics of these models. Drawing on insights from prior studies~\cite{li2023flash,yan2014blogel}, we categorize the eight selected algorithms into three main classes and analyze their suitability across different computing models.

\begin{itemize}[leftmargin=*]
\item \textbf{Iterative Algorithms (\pr, \lpa)} involve multiple iterations, where each vertex is updated based on its neighbors' statuses, continuing until convergence or a fixed number of iterations is reached. The \vc model is well-suited for iterative algorithms as it applies the update function directly to each vertex. These algorithms can also be efficiently implemented using the \ec model, particularly for handling highly skewed graphs. 

\item \textbf{Sequential Algorithms (\sssp, \wcc, \bc, \cd)} require a specific execution order and often involve frequent synchronizations, such as iterations in \sssp. The \blc model is particularly well-suited for such algorithms, especially those with fewer iterations. While it is possible to implement algorithms using the \vc and \ec models, this can lead to suboptimal performance. However, some \vc-based platforms~\cite{10184838, DBLP:conf/ppopp/ShunB13} can execute computation on a vertex subset instead of activating all vertices to minimize unnecessary computations.

\item \textbf{Subgraph Algorithms (\tc, \kc)} involve iterative subgraph matching starting from a set of vertices, requiring intensive communication and computation. The \vc model struggles with handling the high communication overhead in such cases. While the \ec model can do simpler tasks like \tc, where only one edge and its two endpoints are needed to count triangles, it becomes inadequate for more complex subgraphs, such as cliques. In contrast, the \sgc model is specifically designed for graph mining problems, enabling the generation of multiple subgraphs and the definition of computation functions directly based on subgraphs, \mbox{making it well-suited for these tasks.}
\end{itemize}
\vspace{-0.25em}
}

\delete{
\begin{table}[!htbp]
\caption{Topics of graph algorithms.}
\vspace{-1em}
\label{tab:all_algo}
\begin{tabular}{|l|l|}
\hline
\textbf{Topic}                                                         & \textbf{Algorithms}                                                                                                                       \\ \hline \hline
Centrality                                                    & \begin{tabular}[c]{@{}l@{}}PageRank, Personalized PageRank,\\ Betweenness Centrality, Closeness Centrality\end{tabular}          \\ \hline
\begin{tabular}[c]{@{}l@{}}Community\\ Detection\end{tabular} & \begin{tabular}[c]{@{}l@{}}Louvain, Label Propagation,\\ Connected Components\end{tabular}                                      \\ \hline
Similarity                                                    & Jaccard Similarity, Cosine Similarity, SimRank                                                                                   \\ \hline
\begin{tabular}[c]{@{}l@{}}Cohesive\\ Subgraph\end{tabular}   & $k$-Core, $k$-Truss, Maximal Clique                                                                                              \\ \hline
Traversal                                                     & \begin{tabular}[c]{@{}l@{}}BFS, Single Source Shortest Path, \\ Maximum Flow, Minimum Spanning Tree, \\Cycle Detection\end{tabular} \\ \hline
\begin{tabular}[c]{@{}l@{}}Pattern\\ Matching\end{tabular}    & \begin{tabular}[c]{@{}l@{}}Triangle Counting, $k$-Clique,\\ Subgraph Matching, Subgraph Mining\end{tabular}                      \\ \hline
Covering                                                      & \begin{tabular}[c]{@{}l@{}}Minimum Vertex Covering,\\ Maximum Matching, Graph Coloring\end{tabular}                              \\ \hline
\end{tabular}
\end{table}
}

\section{Data Generator \& Synthetic Datasets}
\label{sec:generator}

\begin{figure}[t]\centering
	\scalebox{0.8}[0.8]{\includegraphics{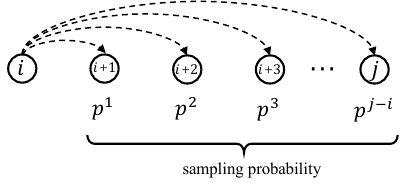}}
\vspace{-1em}
	\caption{Probabilities distribution used in \ldbcdg}
	\label{fig:hop_generator_aa}
\vspace{-0.5em}
\end{figure}

Benchmarking large-scale platforms requires datasets with diverse statistics. However, real-world datasets are often limited due to high collection costs and conflict-of-interest concerns. As a result, synthetic graph datasets have emerged as a practical alternative, generated by data generators that take specific parameters to produce the desired graphs.

Among various data generators, the LDBC Graphalytics generator (\ldbcdg) is noted for its high-quality simulation of real-life social graphs~\cite{prat2014community}. Specifically,
\notation{B.(1)} \revise{\ldbcdg (1) initially generates vertices with some properties (such as \textit{location} and \textit{interest}). (2) Then, vertices are sorted with some specific similarity metrics (e.g., Z-order for \textit{location} and identifier order for \textit{interest}). (3) Finally, edges are formed with probabilities influenced by the distances between vertices in this ordered sequence~\cite{erling2015ldbc}.}

Briefly, \ldbcdg follows the ``Homophily Principle''~\cite{mcpherson2001birds} that individuals with similar characteristics are more likely to form connections. Therefore, similar vertices are closer in distance and have a higher probability of being connected.
\ldbcdg defines the edge formation probability between vertex $u_i$ and $u_j$ as follows:
	$\Pr[e(u_i, u_j)] = max\{p^{|j - i|}, p_{limit}\}$
, where $p$ is the base probability, $i$ and $j$ represent the position of $u_i$ and $u_j$ in the ordered sequence, $p_{limit}$ is a lower-bound. By default, $p$ and $p_{limit}$ are $0.95$ and $0.2$, respectively. To generate edges, \ldbcdg iterates over each vertex $u_i$ and progressively samples $u_j (j > i)$ with the probability $\Pr[e(u_i,u_j)]$ as shown in Figure~\ref{fig:hop_generator_aa} until the number of required edges is obtained. 
\mbox{However, \ldbcdg has two key limitations:}

\begin{itemize}[leftmargin=*]
	\item \textbf{Inflexibility:}  \ldbcdg focuses on producing social networks of different scales. However, graphs of the same scale can differ in characteristics like density and diameter, affecting platform performance. For instance, subgraph algorithms are sensitive to graph density, while sequential algorithms can be impacted by diameter. Some platforms offer optimizations for high-density~\cite{yan2015effective,fan2018parallelizing} or large-diameter graphs~\cite{yan2014blogel}. To effectively evaluate platform performance, the data generator should provide the flexibility of creating graphs with varied characteristics.

	\item \textbf{Inefficiency:} \ldbcdg samples each edge successively and individually. However, the exponential function decays rapidly, leading to a low sampling probability and more unsuccessful attempts, which is time-consuming and inefficient especially when generating a sparse graph.


\end{itemize}

\notation{B.(2)} \revise{To overcome these limitations, we propose the Failure-Free Trial Data Generator~(\fftdg). Compared to \ldbcdg, \fftdg can offer greater flexibility to control the density and diameter as well as better similarity to the real-world datasets, while the efficiency is also improved. Utilizing \fftdg, we can create various datasets with differing scales, densities, and diameters, serving as the default synthetic datasets for our benchmark.}

\subsection{Failure-Free Trial Generator}


Our \fftdg also has three steps and the initial two steps are the same as \ldbcdg. In the final edge generation, instead, \fftdg directly extracts existing edges, bypassing intermediate failed trials. This ensures that the number of trials matches the number of generated edges, significantly reducing unnecessary sampling overhead.

\begin{figure}[t]\centering
	\scalebox{0.9}[0.9]{\includegraphics{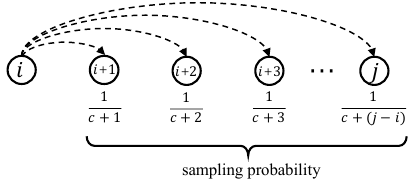}}
\vspace{-1em}
	\caption{Probabilities distribution used in \fftdg}
	\label{fig:hop_generator_a}
\vspace{-0.5em}
\end{figure}

As shown in Figure~\ref{fig:hop_generator_a}, we introduce a new probability function. For a vertex $u_i$, the probability of connecting an edge to $u_j$ $(j > i)$ is defined as $\frac{1}{c + (j - i)}$. $c \ge 0$ is an adjustable parameter, where a larger $c$ can reduce the probability of each edge. In \fftdg, $c$ defaults to $0$ such that the adjacent edge $e(u_i, u_{i+1})$ always exists. \notation{B.(1)} \revise{As verified in Section~\ref{sec:exp_datagen},  this new function does not affect the quality of generated graphs compared to that of \ldbcdg.}

Unlike \ldbcdg, where each edge is sequentially sampled until $e(u_i, u_j)$ exists, \fftdg can directly obtain the first existing edge and avoid sampling intermediate edges from $e(u_i, u_{i + 1})$ to $e(u_i, u_{j-1})$.
Specifically, the probability of an edge $e(u_i, u_j)$ being the first existing edge of $u_i$ is:
\begin{equation}
	\begin{aligned}
		\Pr[e(i, j)] &= \left(1 - \frac{1}{c + 1}\right) \cdot \left(1 - \frac{1}{c + 2}\right) ... \frac{1}{c + (j - i)} \\
		&= \frac{c}{c + (j-i-1)} - \frac{c}{c + (j-i)}
	\end{aligned}
	\label{eq:new_edge_forming}
\end{equation}

Then, the probabilities for each edge to be the first existing edge are structured such that the sequence of probabilities, $1-\frac{c}{c+1}$, $\frac{c}{c+1} - \frac{c}{c+2}$, etc., forms a continuous range that spans from $0$ to $1$.
\notation{B.(1)} \revise{By treating these probabilities as distinct segments that collectively cover the interval $(0, 1]$, we can randomly select a floating-point number $f$ from this interval $(0, 1]$ to represent an edge $e(i, j)$ by}
$
j = i + \left\lfloor \left(\frac{1}{f} - 1 \right) \cdot c \right\rfloor + 1
$.
Once an edge is sampled (e.g., $e(u_i,u_j)$), the following edges $e(u_i,u_k) \ (k {>} j)$ exhibit similar properties:
\begin{equation*}
	\begin{aligned}
		\Pr[e(i, k)] {=} \prod_{t = j + 1 - i}^{k - 1 - i} \left( 1 - \frac{1}{c + t} \right) \cdot \frac{1}{c + (k - i)}
	\end{aligned}
\end{equation*}
Introducing $c' = c + (j - i)$ simplifies the expression:
\begin{equation*}
	\begin{aligned}
		\Pr[e(i, k)]
		= \frac{c'}{c' + (k-j-1)} - \frac{c'}{c' + (k-j)}
	\end{aligned}
\end{equation*}

The probabilities for each edge to become the next existing edge are structured similarly to $1-\frac{c'}{c'+1}$, $\frac{c'}{c'+1} - \frac{c'}{c'+2}$, and so forth.
Consequently, we can consistently obtain connected edges by updating the parameter $c$ without any failure samples. The details are outlined in Algorithm~\ref{algo:distance_hop_generator}.

\begin{algorithm}[t]
	\DontPrintSemicolon
	\small
	\caption{Failure-free Trial Data Generator}
	\label{algo:distance_hop_generator}
	\KwIn{a vertices set $V$, the clustering parameter $\alpha$}
	\KwOut{the edges set $E$}
	\SetKwComment{comment}{$\triangleright$ }{}

	sort $u_i \in V$ with a specific similarity metric

	\For{$u_i \in V$}{
		$c \leftarrow 0, \ j \leftarrow i$

		\While{$u_i.degree < u_i.degree\_limit$}{
			\revise{select $f$ from $(0, 1]$ randomly and uniformly}

			$k \leftarrow j + \left\lfloor \left(\frac{1}{f} - 1 \right) \cdot \frac{c}{\alpha} \right\rfloor + 1$ \comment*{Sampling}


			\textbf{if} $k > \vert V \vert$ \textbf{then Break}

			\If{$u_k$.degree $<u_k$.degree\_limit}{
				$E \leftarrow E \cup e(i, k)$ \comment*{Generating}
			}

			$c \leftarrow c + (k - j), \ j \leftarrow k$
		}
	}
	\Return $E$
\end{algorithm}

\subsection{Flexibility Improvement}
We present adjustments to \fftdg (Algorithm~\ref{algo:distance_hop_generator}) for generating graphs with varying densities and diameters. Note that density enhancement is feasible because of the proposed edge formation probability in Equation~\ref{eq:new_edge_forming}, while diameter adjustment can be integrated smoothly into \ldbcdg.
\subsubsection{Density Enhancement}
The value of the sampling functions both in \ldbcdg and our generator, i.e., $p^{\vert j - i \vert}$ and $\frac{1}{c + \vert j - i \vert}$, decreases rapidly and can only generate few edges.

To enhance the density and generate more edges, \ldbcdg adds a lower bound on the probability $p_{limit}$. However, this resolution is overly simplistic and lacks justification, as it assigns the same connection probability $p_{limit}$ to all distant vertices, regardless of their actual separation.

\stitle{Density Factor.} Our approach is adding a density factor $\alpha > 1$ \revise{(which is used to control the density of the generated graph)} during the sampling process to enhance the generation of edges, as shown in Line 5 of Algorithm~\ref{algo:distance_hop_generator}.
The factor $\alpha$ serves to concentrate probability from distant to local vertices. More precisely, we can reformulate the probability of an edge being the first existing edge as follows:
	$\Pr[e(i,j)] = \frac{1}{c + \frac{d(d-1)}{c} + 2d-1}$,
where $d=j-i$ is the distance. The problem is the change occurs when $c$ is reduced to $\frac{c}{\alpha}$. An observation is that $\Pr[e(i,j)]$ has a pair of the same values when $c = \frac{d(d-1)}{c}$. If $c < \frac{d(d-1)}{c}$, $Pr[e(i, j)]$ is always smaller. Otherwise, only when $\frac{c}{\alpha} \le \frac{d(d-1)}{c}$ can $\Pr[e(i,j)]$ become lower. Combining two conditions, we have:
\begin{equation*}
	\begin{aligned}
		d(d-1) \ge \frac{c^2}{\alpha}, \quad
		d \ge \frac{1+\sqrt{1+\frac{4c^2}{\alpha}}}{2}
	\end{aligned}
\end{equation*}

For convenience, we can use $d \ge \sqrt{c^2 / \alpha}$ to quickly check since $d \approx d-1$ when $d$ is large. Therefore, if $d$ is larger than this boundary, i.e., the vertices are too far away, the probability $\Pr[e(i,j)]$ decreases due to the influence of $\alpha$. Correspondingly, other nearby vertices will have a higher probability.
\notation{B.(1)} \revise{When the value of $\alpha$ is larger, the probability of each new edge being close to the edges generated in the previous round increases, which leads to a higher graph density.
Though it is difficult to quantify the impact of $\alpha$, our experiments show that increasing $\alpha$ ten-fold results in approximately twice as many edges being generated.}

\begin{table}[h]
	\caption{Selected synthetic datasets}
	\label{tab:syn_data}
	\vspace{-0.8em}
	\small
	\begin{tabular}{c|c|c|c|c}
		\hline
		\textbf{Datasets} & \textbf{n} & \textbf{m} & \textbf{Density} & \textbf{Diameter} \\ \hline 
		S8-Std            & 3.6M       & 153M       & $2.4 \times 10^{-5}$     & 6                           \\ \hline
		S8-Dense          & 1.2M       & 159M       & $2.2 \times 10^{-4}$     & 5                           \\ \hline
		S8-Diam           & 3.6M       & 155M       & $2.4 \times 10^{-5}$     & 101                         \\ \hline
		S9-Std            & 27.2M      & 1.42B      & $3.8 \times 10^{-6}$     & 6          \\ \hline
		S9-Dense          & 9.1M       & 1.47B      & $3.6 \times 10^{-5}$     & 5                            \\ \hline
		S9-Diam           & 27.2M      & 1.48B      & $4.0 \times 10^{-6}$     & 102        \\ \hline
		S9.5-Std         & 77M        & 4.36B      & $1.5 \times 10^{-6}$     & 6          \\ \hline
		S10-Std           & 210M       & 12.62B     & $5.7 \times 10^{-7}$     &  6        \\ \hline
	\end{tabular}
\end{table}




\subsubsection{Diameter Adjustment.}
The diameter is a crucial metric, particularly in the context of distributed algorithms, yet it is often overlooked in many synthetic datasets. Thus, we propose a method that helps adjust diameters to accommodate various needs.

\begin{figure}[t!]
	\scalebox{0.75}[0.75]{\includegraphics{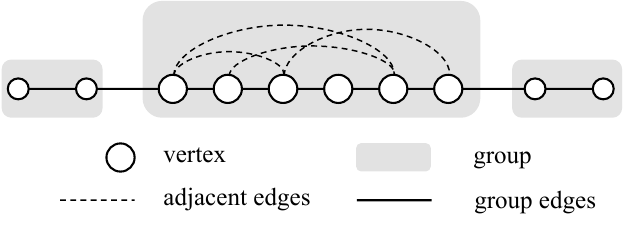}}
\vspace{-1em}
	\caption{Generating graphs with adjustable diameters}
	\label{fig:large_diameter}
\vspace{-1em}
\end{figure}

Our approach restricts that most edges are within a group. As depicted in Figure~\ref{fig:large_diameter}, all vertices are organized into distinct groups.
Then, edges are classified into two types: adjacent edges and group edges. Initially, adjacent edges are established between neighboring vertices to guarantee connectivity.
Following this, group edges within each group are generated using \fftdg. The diameter within each group remains relatively constant and independent of the group's size, averaging about $6$, thus allowing control over the number of groups to produce a range of diameters:
\begin{equation*}
	\begin{aligned}
		group\_number = \frac{target\_diameter}{group\_diameter + 1},
		group\_size = \frac{\vert V \vert}{group\_number}
	\end{aligned}
\end{equation*}
The adjustment involves an additional check in Line~7 of Algorithm~\ref{algo:distance_hop_generator} that the iteration also breaks once
the vertices with indices $i$ and $k$ do not belong to the same group as:
$
    \left\lfloor \frac{i}{group\_size} \right\rfloor \ne \left\lfloor \frac{k}{group\_size} \right\rfloor
$.
Once $k$ is too large to exceed the group, the generation process for $u_i$ is finished.

\subsection{Synthetic Datasets}
\label{sec:selected_datasets}

Based on \fftdg, we create eight synthetic datasets with differing scales, densities, and diameters, serving as the default synthetic datasets for our benchmark. Table~\ref{tab:syn_data} shows their details.

Specifically, we design four different scales, where the vertex numbers are correlated to the LDBC default setting, i.e., $3.6$M, $27.2$M, $77$M, and $210$M. We also provide two datasets with an enhanced density~(i.e., $1.2$M and $9.1$M vertices and the same number of edges) and two datasets with a larger \mbox{diameter~(i.e., $\sim 100$ with the same scale).}

Each dataset is named based on its size scale and characteristic features. The size scale is quantified as the base-$10$ logarithm of the sum of the number of vertices and edges, denoted as $\log_{10}(\vert V \vert + \vert E \vert)$. The suffix \textit{Std}, \textit{Dense} and \textit{Diam} represent standard social networks, dense networks and large-diameter networks, respectively.

\section{Benchmarking API Usability}
\label{sec:evaluator} 


\subsection{An Overlooked Benchmarking Scope}


The API usability is a qualitative characteristic that evaluates how easy it is to learn and use~\cite{DBLP:journals/csr/RaufTP19,DBLP:journals/cacm/MyersS16}. Highly usable APIs enhance productivity and reduce bugs, while initial interactions with poorly designed APIs can leave a lasting impression of complexity, potentially deterring users from embracing the platform. 
The most common evaluation method is the empirical study, where programmers are recruited to answer interview questions~\cite{farooq2010api,piccioni2013empirical}. However, this approach is costly and not scalable, as recruiting participants with adequate programming skills is challenging, and the number of possible usage scenarios for an API can be vast~\cite{farooq2010api}. 
Consequently, most graph analytics platform benchmarks overlook API usability evaluations, instead reporting the lines of code required for specific tasks as an indicator of API usability~\cite{6976096,DBLP:conf/kbse/NamHMMV19,watson2009improving}. Clearly, this approach is oversimplified and inadequate.

To address the shortcomings in existing API usability benchmarks, we propose a scalable, automated approach.
Our idea is that since these shortcomings are primarily due to human factors, and LLMs have demonstrated the ability to match or surpass human abilities in many areas, why not use LLMs to replace humans in API usability evaluation? 
This approach eliminates the need for qualified programmers, cuts costs, and enables large-scale evaluation.

\begin{figure}[t]
	\begin{center}
		\begin{tabular}[t]{c}
			\includegraphics[width=0.95\columnwidth]{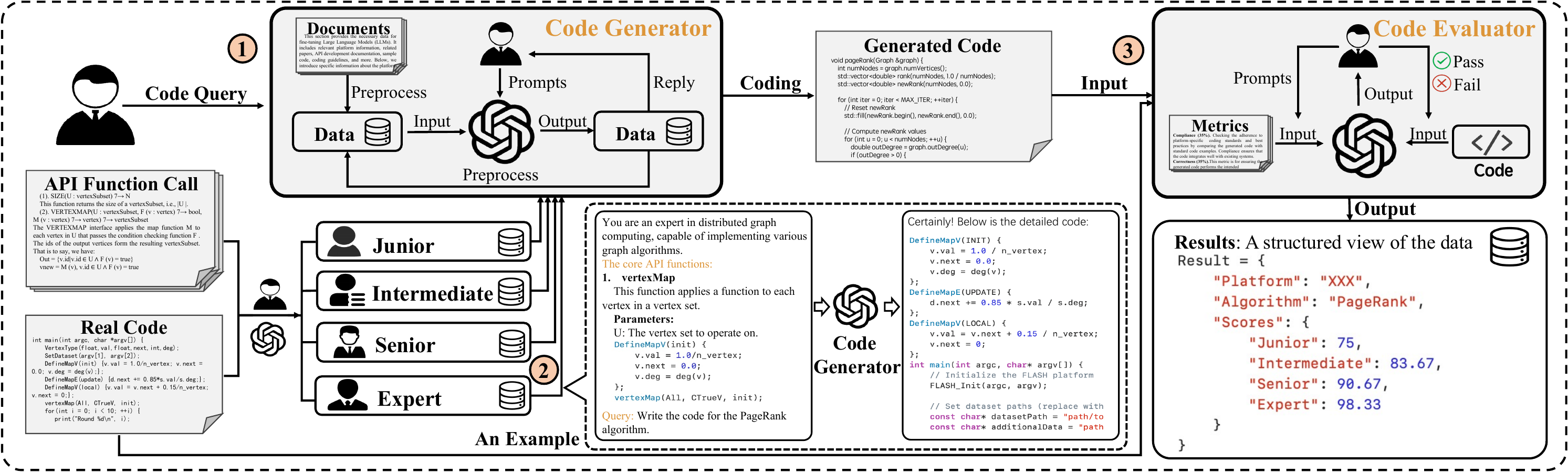}
		\end{tabular}
	\end{center}
	 \vspace{-1em}
	\caption{LLM-based usability evaluation framework}
	\label{fig:llm_framework}
	 \vspace{-0.5em}
\end{figure}



\stitle{Our Solutions.} To ensure LLMs familiarize each platform, we perform instruction-tuning using a comprehensive dataset, including platform-specific API documentation, research papers, and sample code. This enhances their ability to generate accurate code. 
However, determining the extent of the LLM's proficiency with different platforms is still challenging. To address this, we use multi-level prompts (from junior to expert) to simulate varying skill levels and stabilize the LLM's understanding.  
We also evaluate the generated code by comparing it with standard reference code and assessing its correctness and readability, ensuring effective use of platform-specific APIs.
Our proposed multi-level LLM-based usability evaluation \mbox{framework is shown in Figure~\ref{fig:llm_framework} and detailed in Section~\ref{sec:llm-based_subjective_evaluation}.}

\subsection{A Multi-Level LLM-Based Usability Evaluation Framework}
\label{sec:llm-based_subjective_evaluation}

\stitle{Overview of the Framework.} Our idea is to simulate programmers with varying skill levels by using LLMs with multi-level prompts. 
By analyzing the quality of the generated code, we can evaluate the usability of platform APIs. 
Specifically, we (1) perform instruction-tuning on LLMs to train the \emph{Code Generator} for each platform by using their public documentation; (2) use the \emph{Code Generator} to produce the code we requested by providing multi-level prompts; and (3) employ the instruction-tuned LLM-based \emph{Code Evaluator} to score the generated code.


\begin{figure}\centering
	\scalebox{0.37}[0.37]{\includegraphics{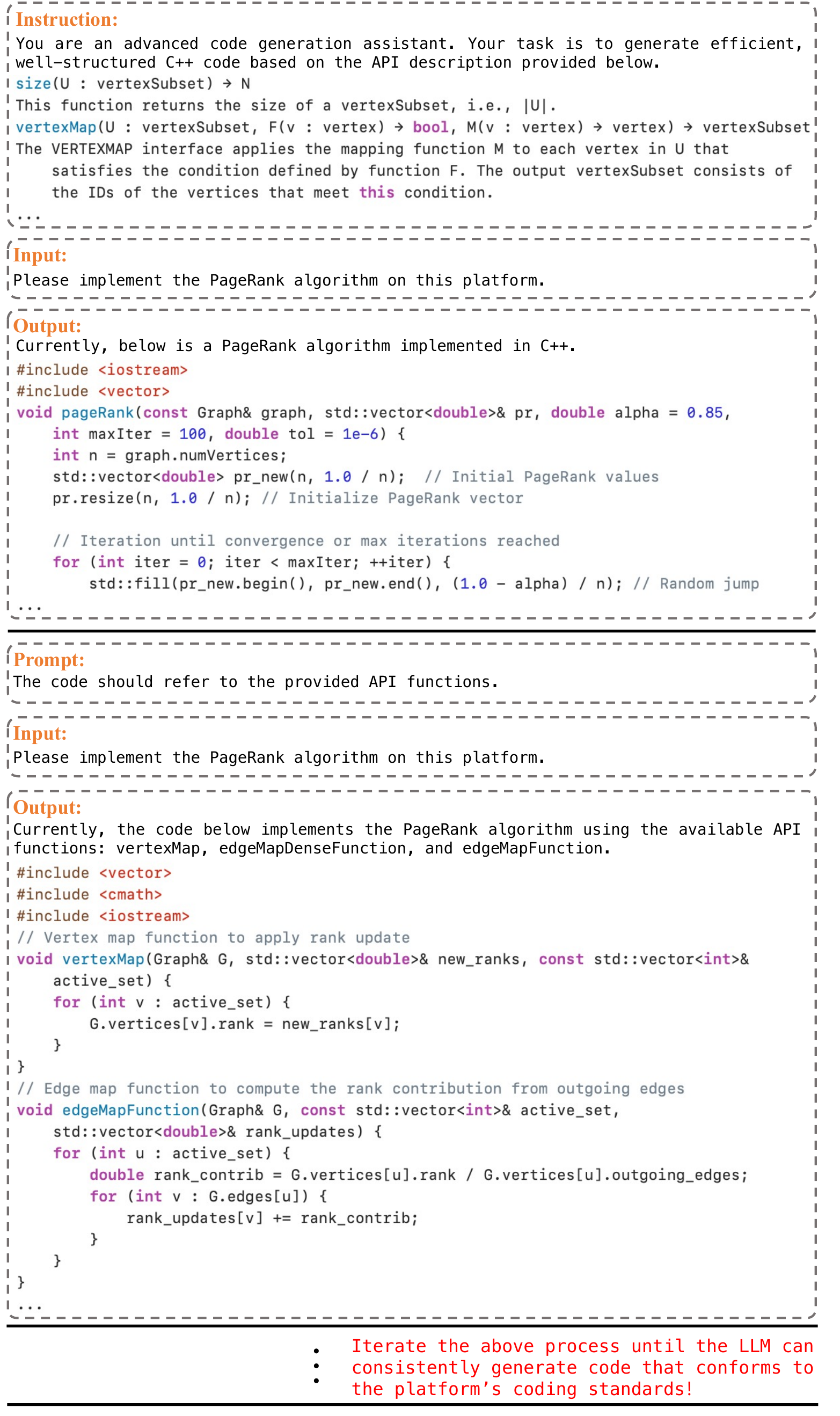}}
	\vspace{-1em}
	\caption{An example of the instruction-tuning process for the code generator}
	\label{fig:llm_instruction_tuning_example}
	\vspace{-2em}
\end{figure}

\stitle{\underline{Step 1: Instruction-Tuning of LLMs.}} Step 1 aims to enable LLMs to generate code aligned with each platform's style, standards, and best practices, particularly in the use of platform-specific APIs, by learning from publicly available documents.
As shown in Figure~\ref{fig:llm_framework}~\textcircled{1}, we gather and preprocess data from each platform, including research papers, API documentation, sample code, and coding guidelines, enabling LLMs to learn specific coding styles and conventions. Human evaluators then review the generated results, providing feedback and prompts, which refine the data and the instruction-tuning process. This iterative process continues until the LLM consistently generates code that complies with platform standards.

Figure~\ref{fig:llm_instruction_tuning_example} illustrates how the LLM is trained to generate PageRank code on a specific platform. 
The example shows the evolution of the code, from general C++ to platform-specific API usage. The LLM is given instructions and prompts to adjust its understanding of platform-specific functions, refining the generated code until it aligns with the platform's best practices, such as using functions like \texttt{vertexMap} and \texttt{edgeMapFunction}.

To ensure fairness and avoid bias during the code generation and evaluation process, all platform-specific identifiers in the API functions are anonymized. During the instruction-tuning process, platform-specific references, such as unique function names or parameters, are modified. This guarantees that LLMs evaluate the usability of APIs based on their general usability characteristics rather than familiarity with specific platforms.

\begin{table*}[t]
	\centering
	\caption{Performance evaluation metrics}
	\vspace{-0.8em}
	\label{tab:performance_evaluation_metrics}
	\small
	\begin{tabularx}{\textwidth}{c|>{\bfseries}c|X}
		\hline
		\textbf{Category} & \textbf{Metric} & \textbf{Description} \\ \hline 
		
		\multirow{5}{*}{\textbf{Timing}} & \multirow{2}{*}{Upload Time} & Time required to read, convert, partition, and load graph data into memory. \\ \cline{2-3}
		& Running Time & Total time required to complete an algorithm execution task. \\ \cline{2-3}
		& \multirow{2}{*}{Makespan} & Overall time for graph operations, including reading, processing, and writing data. \\ \hline
		
		\textbf{Throughput} & Edges/sec & Number of edges processed per second. \\ \hline

		\multirow{2}{*}{\textbf{Scalability}} & \multirow{2}{*}{Speedup} & Rate of performance improvement with additional computational resources. \\ \hline
		
		\textbf{Robustness} & Stress Test & Platform's stability and reliability under high-stress conditions. \\ \hline
		
	\end{tabularx}
	\vspace{-0.5em}
\end{table*}

\stitle{\underline{Step 2: Multi-Level Prompts.}} 
Step 2 simulates programmers with varying expertise levels: junior, intermediate, senior, and expert, as shown in Figure \ref{fig:llm_framework} \textcircled{2}.
The \emph{Code Generator} uses varying prompts to generate corresponding code, and we evaluate the difficulty of using the API by examining the generated code's quality, thereby assessing the platform's API usability.
For example, high scores at lower-level prompts indicate ease of use and low learning costs of APIs on this platform. The prompt levels and their respective details are as follows:


\begin{itemize} [leftmargin=*]
	\item \textbf{Level 1 (Junior).} At this level, no specific technical details are provided. The prompt only consists of a description of the task, without any guidance on how to implement it.
	
	\item \textbf{Level 2 (Intermediate).} This level offers minimal technical information to guide the code generation. Basic prompts are given, including the names of core APIs and parameters.
	
	\item \textbf{Level 3 (Senior).} This level provides detailed usage instructions for the relevant APIs, including the names of the APIs and parameters, and a detailed introduction to them. Besides, some example code is provided to guide the usage of API functions. 
	
	\item \textbf{Level 4 (Expert).} In addition to the detailed API instructions similar to the previous level, this level also includes the pseudo-code of the relevant algorithm.
\end{itemize}

\revise{To ensure fairness in our usability evaluation, we strictly rely on the most fundamental (i.e., lowest-level) APIs provided by each platform, such as \texttt{compute()} and \texttt{reducer()} in \pregel or \texttt{gather()}, \texttt{apply()}, and \texttt{scatter()} in \power. This approach ensures that the evaluation reflects the flexibility and extensibility of the API design itself rather than relying on simplified, high-level wrappers. }


\stitle{\underline{Step 3: Code Evaluation.}} 
Step 3 is crucial for determining the quality of the generated code and the API's usability. 
We first define the evaluation metrics. Correctness and readability are commonly used metrics in API usability evaluation~\cite{DBLP:journals/csr/RaufTP19,DBLP:conf/icse/Myers17,DBLP:conf/esem/PiccioniFM13}. Correct code indicates that the API is intuitive and that the provided documentation and examples effectively guide developers in implementing the desired functionality correctly. 
High readability suggests that the API's design promotes clear and concise coding practices. These two metrics are crucial for assessing usability. 
\notation{C.(3)}\revise{However, we observe that LLMs often exhibit ``hallucination'', focusing on general programming patterns or inventing nonexistent API functions while ignoring platform-specific APIs, typically due to unclear or ambiguous prompts. This limitation mirrors the behavior of human developers: less experienced programmers are more likely to make mistakes when dealing with poorly designed APIs. 
To address this, we introduce a new metric, compliance, which measures how closely the generated code aligns with standard code. This metric reflects the API's intuitiveness and accessibility for developers with varying skill levels. By assessing whether the API enables users to easily produce code that establishes best practices and standards, compliance provides an objective basis for scoring API usability.}
The details of evaluation metrics and their weights are as follows:

\begin{itemize}[leftmargin=*]
	\item \textbf{Compliance (35\%).}  Checking the adherence to platform-specific coding standards and best practices by comparing the generated code with standard code examples. Compliance ensures that the code integrates well with existing systems.
	
	\item \textbf{Correctness (35\%).} This metric is for ensuring the generated code performs the intended task accurately. This includes verifying the logic of the code and the correctness of function calls.
	
	\item \textbf{Readability (30\%).} This metric focuses on code clarity and maintainability. Readable code is easier to understand, modify, and debug. It should be well-structured, logically grouped, and follow consistent naming conventions.
\end{itemize}

\revise{
A simple method would be to assign equal weights, but this leads to an indivisibility issue.
Considering that compliance and correctness more directly reflect the LLM's understanding of platform APIs compared to readability, we assign slightly higher weights to compliance and correctness.
Users can also customize the weight distribution based on their specific needs.}
\revise{During evaluation, we provide \emph{Code Evaluator} with the evaluated and standard code and ask it to generate scores for each metric based on our provided guidelines.
Figure~\ref{fig:llm_framework}\textcircled{3} illustrates the training process: we first provide detailed scoring criteria as the LLM's knowledge base and set basic requirements instructions to get \emph{Code Evaluator}. Then, we introduce some test codes and provide feedback based on the output results to optimize the Code Evaluator's instructions. We iterate this process until the LLM can produce stable and satisfactory evaluation results.}


\section{\revise{Overall Process of Our Benchmark}}
\label{sec:performance_evaluation_metrics}

\begin{figure}[t]\centering
	\scalebox{0.25}[0.25]{\includegraphics{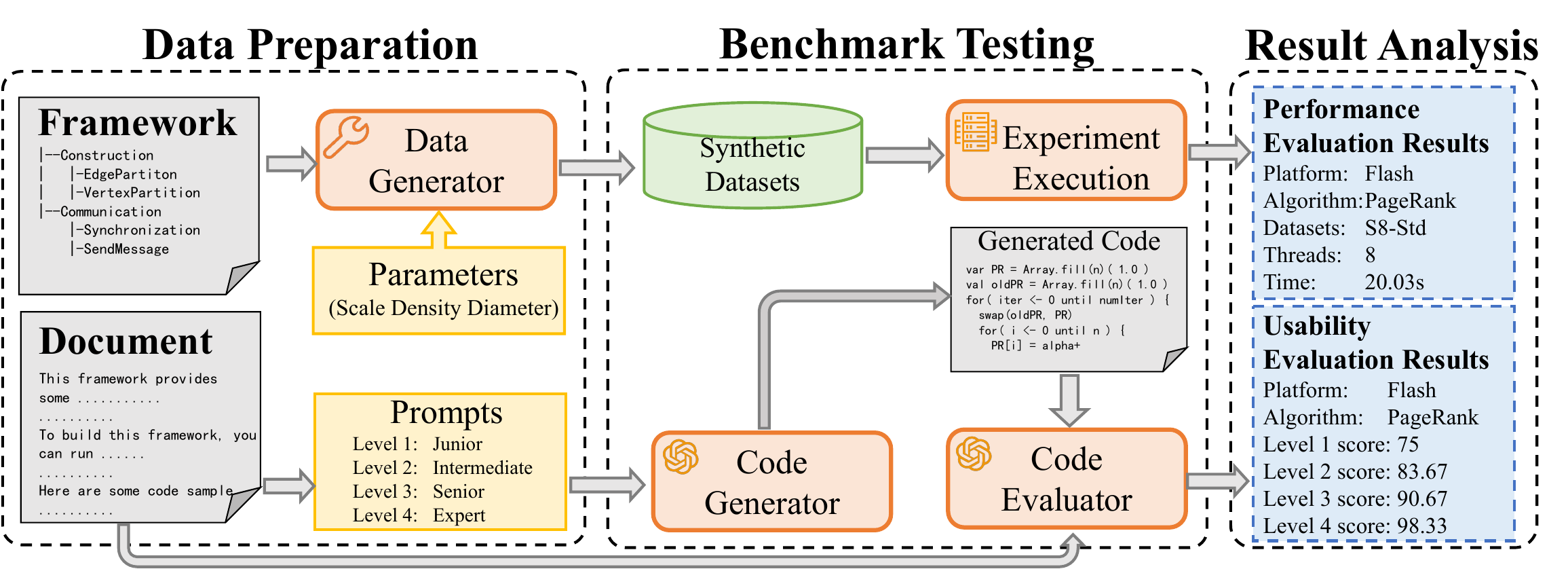}}
	\vspace{-0.8em}
	\caption{An overview of our benchmark}
	\label{fig:intro_overview}
	 \vspace{-0.8em}
\end{figure}

\revise{In this section, we provide a detailed overview of our benchmark's workflow, aiming to offer an in-depth evaluation of platforms from the perspectives of both performance and usability.}
As illustrated in Figure~\ref{fig:intro_overview}, the architecture of our benchmark includes three main stages: data preparation, benchmark testing, and result analysis.

\stitle{\underline{Data Preparation.}} 
In this stage, we generate and preprocess data for benchmarking. Using a flexible \textit{Data Generator} (Section~\ref{sec:generator}), we create synthetic graph datasets by varying parameters (e.g., scale, density, diameter) and convert them into platform-compatible formats. 
We also gather and preprocess platform resources—research papers, API documentation, and sample code—to meet the requirements of our usability evaluation framework (Section~\ref{sec:evaluator}). Different prompts targeting platform-specific algorithms are prepared for the \textit{Code Generator}, and to avoid bias, all platform-related information is concealed in both prompts and platform data.

\stitle{\underline{Benchmark Testing.}} 
In this stage, we evaluate both performance and API usability. Using the \textit{Experiment Executor}, we run selected core algorithms on synthetic datasets across various platforms. We measure metrics such as timing, throughput, scalability, and robustness (summarized in Table~\ref{tab:performance_evaluation_metrics}) to gauge each platform-algorithm combination’s efficiency and effectiveness.

In the API usability evaluation stage, the \textit{Code Generator} uses predefined prompts to produce code, which is combined with the platform-provided code and scored by the \textit{Code Evaluator}. To reduce bias, we repeat this process multiple times and average the results to obtain the final score.

\stitle{\underline{Result Analysis.}} 
In this stage, we aggregate and analyze both performance outcomes and API usability scores to reveal each platform’s strengths and weaknesses. An example of the experimental results analysis is provided in Section~\ref{sec:guide}.

\delete{
	Figure~\ref{fig:intro_overview} presents a comprehensive illustration of our benchmark architecture, which consists of four primary components:
	\begin{itemize}[leftmargin=*]
		\item \textbf{Data Generator} receives a variety of parameters~(e.g., scale, density, diameter, etc.) to generate synthetic datasets with diverse statistical properties. The generator also has flexible storage, adapting to various output schemas required by different platforms.
		
		\item \textbf{Experiment Executor} executes selected algorithms on all platforms using the generated datasets. The executor meticulously measures and records performance using a comprehensive set of evaluation metrics, which are categorized into four key areas, including timing, throughput, scalability, and robustness. The detailed descriptions of these metrics are summarized in Table~\ref{tab:performance_evaluation_metrics}. These metrics provide in-depth insights into the efficiency and effectiveness of each platform-algorithm combination.
		
		\item \textbf{Code Generator} collects and analyzes public documents of platforms of each platform, utilizing LLM to generate corresponding code using different levels of prompts. The prompts have four levels including Junior, Intermediate, Senior, and Expert, simulating different degrees of user expertise.
		
		\item \textbf{Code Evaluator} performs a comprehensive comparison between the LLM-generated code and standardized, expert-validated code. The evaluation employs a range of evaluation criteria including compliance, correctness and readability, ultimately producing a comprehensive final score that reflects the quality and usability of the generated code.
	\end{itemize}
	
	These four components form two individual evaluation processes within our benchmark, namely Performance Evaluation and Usability Evaluation. The Performance Evaluation begins by employing the Data Generator to generate a series of synthetics, which are then processed by the Experiment Executor to run code. Concurrently, the Usability Evaluation utilizes the Code Generator to produce LLM-generated code, which is subsequently compared with standard code and scored by the Code Evaluator. Upon completion of these evaluations, the performance evaluation result and the usability evaluation result are separately reported, providing a holistic view of a platform's efficiency and user-friendliness.
}

\section{Experimental Setup}
\label{sec:exp_setup}
\subsection{Environment}

We evaluate \revise{seven} widely used graph analytics platforms. Each platform employs distinct programming languages, storage formats, communication methods, and \revise{computation models}, which are summarized in Table~\ref{tab:platform_model}.

\begin{table}[t]
  \small
  
  \caption{\revise{Programming languages and computing models (Section \ref{sec:algorithm_suitability}) used in each platform \notation{A.(3)}}}
  \vspace{-0.8em}
  \label{tab:platform_model}
\begin{tabular}{c|c|c}
  \hline
  \textbf{Platform }  & \textbf{Language}   & \textbf{Execution Model}       \\ \hline
  \graphx    & Scala      & \vc          \\ \hline
  \power     & C++        & \ec              \\ \hline
  \flash     & C++        & \vc        \\ \hline
  \grape     & C++ / Java & \blc                  \\ \hline
  \pregel    & C++        & \vc           \\ \hline
  \ligra     & C++        & \vc       \\ \hline
  \gthinker  & C++        & \sgc                  \\ \hline
  \end{tabular}
  \vspace{-0.5em}
\end{table}

\vspace{-0.2em}
\begin{itemize}[leftmargin=*]

\item \textbf{GraphX (GX)}~\cite{gonzalez2014graphx} is a component in Spark adapted for graph-parallel computation, \revise{which provides the \vc model using Spark Resilient Distributed Datasets (RDD) APIs}.

\item\textbf{PowerGraph (PG)}~\cite{Low+al:uai10graphlab} is an \revise{\ec} distributed graph analysis platform, addressing the difficulties including load imbalance in \mbox{graph algorithms over real-world power-law graphs.} 

\item\textbf{Flash (FL)}~\cite{10184838} is a distributed platform, supporting various complex graph algorithms including clustering, centrality, traversal, etc. It \revise{extends
the \vc model with global vertex status to make it easy to program complex graph algorithms}.  

\item\textbf{\grape (GR)}~\cite{10.1145/3035918.3035942} is a \revise{\blc parallel platform, which is designed to execute existing (textbook) sequential graph algorithms with only minor modifications in the distribution context}.

\item\textbf{Pregel+ (PP)}~\cite{yan2015effective} 
\revise{extends the original \vc platform Pregel~\cite{malewicz2010pregel} by introducing vertex mirroring and efficient message reduction, making it effective for power-law and dense graphs.} 

\item\textbf{Ligra (LI)}~\cite{DBLP:conf/ppopp/ShunB13} is a lightweight \revise{\vc} processing platform for shared memory on a single machine with multiple cores. Ligra is applicable if one machine can allocate the whole graph. 

\revise{\item\textbf{G-thinker (GT)}~\cite{yan2020g} is a \sgc platform specialized for graph mining problems, which can create and schedule subgraphs with arbitrary shapes and sizes.}
\end{itemize}
\vspace{-0.2em}

All experiments are conducted on a cluster with $16$ machines. Each machine has $4$ Intel$^\circledR$  Xeon$^\circledR$ Platinum 8163 @ 2.50GHz CPUs, 512 GB memory, and 3 TB disk space. The 16 machines are connected with a 15  Gbps LAN network.


\begin{table*}[t]\centering
    \caption{Experimental methodology}
    \vspace{-0.8em}
    \label{tab:methodology}
    \resizebox{\textwidth}{!}{
\begin{tabular}{>{\centering\arraybackslash}p{3cm}|>{\centering\arraybackslash}p{2.5cm}|>{\centering\arraybackslash}p{2.5cm}|>{\centering\arraybackslash}p{3.8cm}|>{\centering\arraybackslash}p{2.2cm}|>{\centering\arraybackslash}p{1.6cm}}
    \hline
    \textbf{Aspects}   & \textbf{Section}   & \textbf{Algorithms}   & \textbf{Datasets}           & \textbf{\#threads}          & \textbf{\#machines}         \\ \hline
    Algorithm Impact       & \multirow{2}{*}{Section~\ref{sec:7.1}} & \multirow{2}{*}{All}          & \multirow{2}{*}{S8-Std, S8-Dense, S8-Diam}                                                                     & \multirow{2}{*}{32} & \multirow{2}{*}{1} \\ \cline{1-1}
    Statistics Impact       &                              &                               &                                                                                                                &                                     &                    \\ \hline
    \multirow{2}{*}{Scalability Sensitivity}   & \multirow{2}{*}{Section~\ref{sec:7.2}} & \multirow{2}{*}{\pr, \sssp, \tc} & S8-Std, S8-Dense, S8-Diam & 1, 2, 4, 8, 16, 32 & 1                  \\  \cline{4-6}
     &                              &                               &                         S9-Std, S9-Dense, S9-Diam                    &             32                      & 1, 2, 4, 8, 16     \\ \hline
     \multirow{2}{*}{Throughput}   & \multirow{2}{*}{Appendix~\cite{gab}} & \multirow{2}{*}{\pr, \sssp, \tc} & \multirow{2}{*}{\begin{tabular}[c]{@{}c@{}}S8-Std, S8-Dense, S8-Diam,\\ S9-Std, S9-Dense, S9-Diam\end{tabular}} & \multirow{2}{*}{32} & \multirow{2}{*}{16}                  \\
     &                              &                               &                                                                                                                &                                     &     \\ \hline
      \multirow{2}{*}{Stress Test}    & \multirow{2}{*}{Appendix~\cite{gab}} & \multirow{2}{*}{\pr} & S8-Std, S9-Std, S9.5-Std,  S10-Std  & \multirow{2}{*}{32} & \multirow{2}{*}{16} \\ \hline
    Usability Evaluation    & Section~\ref{sec:7.5} & All & ---  & --- & --- \\ \hline

     \end{tabular}
     }
     \vspace{-1em}
\end{table*}

\subsection{Algorithms}
We utilize all eight algorithms outlined in Section~\ref{sec:selected_algorithms} to evaluate the platforms, with the following settings: For PageRank (\pr) and Label Propagation Algorithm (\lpa), the maximum number of iterations is set to 10. For Triangle Counting (\tc) and $k$-Clique (\clique), only the counting results are reported. In Single Source Shortest Path (\sssp) and Betweenness Centrality (\bc), the source vertex is set to 0. For Connected Components (\cc), we use undirected networks (i.e., weakly connected) to identify all connected components. In \core, the minimum coreness value starts at 1 and is gradually increased until all vertices are removed. Detailed implementation descriptions for all algorithms can be found in the artifact.


\delete{
\begin{itemize}[leftmargin=*]
\item\textbf{PageRank.} In each iteration, each vertex collects old PageRank values from neighbors to compute its new PageRank value. The maximum iteration number is $10$.

\item\textbf{SSSP.} In each iteration, each vertex collects distance values from neighbors to update its new distance value until convergence, i.e., no distance value is updated. If the platform only supports unweighted graphs, we assume all edge weights are the same. The source vertex is $0$.

\item\textbf{Triangle Counting.} For each edge, collect the neighbors of two endpoints and compute the intersection. The triangle counting number is the summation of all intersection sizes.

\item\textbf{Connected Component.} Initially, each vertex has a unique ID. In each iteration, each vertex finds the minimum IDs from its neighbors as its new ID until convergence. All vertices with the same ID form a connected component.

\item\textbf{Betweenness Centrality.} We compute the betweenness centrality on the unweighted graph and apply the Brandes[cite] algorithm, which has a BFS process and a backtrace process to compute the shortest path number and the partial betweenness centrality value, respectively. The BFS root is the vertex $0$.

\item\textbf{Label Propagation Algorithm.} Initially, each vertex has a unique label. In each iteration, each vertex selects the most frequent label from its neighbors as its new label.

\item\textbf{Core Decomposition.} Intially $k=1$ and gradually enlarge $k$. For each $k$, repeatedly deleting vertices with degrees lower than $k$. Finally, the algorithm converges when all vertices are deleted.

\item\textbf{k-Clique.} Initially, each vertex sets its neighbors as the candidate set. Then send candidate set message to the candidate set vertices, intersecting its neighbors as a new candidate set. After $k-1$ times intersection, previous $k-1$ vertices with any vertex in the candidate set can form a k-Clique. The $k$ value is $5$.

\end{itemize}
}


\subsection{Datasets}
We test seven synthetic datasets, as detailed in Section~\ref{sec:selected_datasets}. Note that evaluations on real-life datasets are similar to the LDBC benchmark~\cite{DBLP:journals/pvldb/IosupHNHPMCCSAT16} and are therefore not the focus of this work. As mentioned above, these datasets are in four scales: S8, S9, S9.5, and S10, with S10 being the largest one. Additionally, there are three variants for S8 and S9: Std, Dense, and Diam. Here, Std represents the standard dataset ($\alpha$=10), Dense denotes a denser dataset ($\alpha$=1000), and Diam indicates a dataset with a larger diameter ($\alpha$=10).

\vspace{-0.5em}
\subsection{Methodology}


Our evaluation methodology, detailed in Table~\ref{tab:methodology}, encompasses six aspects. Unless stated otherwise, the experiments are conducted using three algorithms: \pr, \sssp, and \tc. These algorithms represent the Iterative, Sequential, and Subgraph, respectively, as discussed in Section~\ref{sec:algorithm_suitability}.

\vspace{-0.4em}
\begin{itemize}[leftmargin=*]
\item \textbf{Algorithm Impact} evaluates a platform's ability to support different graph algorithms. We use 32 threads on a single machine to run all algorithms on three S8 datasets with varying statistics.

\item \textbf{Statistics Impact} assesses platform sensitivity to various dataset characteristics by running all algorithms on the three S8 datasets. This experiment highlights platform optimizations that enable effective handling of datasets with diverse characteristics, which may not be evident when using a single dataset alone.

\item \textbf{Scalability Sensitivity} assesses both scale-up and scale-out performance. For scale-up, we increase the number of threads on a single machine (1, 2, 4, 8, 16, 32). For scale-out, we increase the number of machines (1, 2, 4, 8, 16), each utilizing 32 threads. To evaluate the platform's sensitivity to different datasets, we use datasets with varying statistics, including the larger S9 datasets while testing scale-out performance.

\item\textbf{Throughput} measures data processing capacity. We run the default algorithms on $S8$ and $S9$ across $16$ machines, each configured with $32$ threads.

\item \textbf{Stress Test} aims to determine the largest dataset that each platform can handle on 16 machines. To achieve this, we run the PageRank (PR) algorithm on datasets S8-Std, S9-Std, S9.5-Std, and S10-Std until the platform is unable to process the dataset, revealing its limitations.

\item \textbf{Usability Evaluation} involves the pioneering subjective assessment of the platform's API usability using our LLM-based evaluation framework, as described in Section~\ref{sec:llm-based_subjective_evaluation}. This evaluation is powered by the GPT-4o model~\cite{GPT-4o}. To ensure the evaluation is convincing, we limit our evaluation to code provided by the platform and the original authors, excluding any code we have written ourselves.

\end{itemize}
\vspace{-0.4em}

\notation{D.(3)}\revise{The TPC-H price/performance is also an important metric.
However, in our experiments, all systems are deployed on the same cluster with 16 machines, ensuring the cost remains constant. Moreover, the processing time and throughput evaluated in the paper partially reflect the system's performance.
Thus, the throughput indirectly reflects the price/performance metric, where a higher throughput indicates a better price/performance ratio.}


\section{Experimental Result}
\label{sec:exp_result}

\subsection{Generation Similarity and Efficiency}
\label{sec:exp_datagen}

\stitle{Generation Similarity.} To illustrate better alignment with real-world graph characteristics, we follow the previous evaluation methodology~\cite{prat2014community}, i.e., generate communities over the social network and present the distribution of community statistics in the dataset. We choose the LiveJournal~\cite{snapnets} dataset as the ground truth and select six statistics, Clustering Coefficient, Triangle Participation Ratio, Bridge Ratio, Diameter, Conductance, and Size. We use the Jensen-Shannon Divergence to evaluate the similarity and the result in Table~\ref{tab:JS} suggests that our datasets achieve $2$x lower divergence on average, which verifies that \fftdg can generate social networks that follow the real-world distribution and are similar to real-world datasets. Figure~\ref{fig:generator} presents the distribution of three major statistics with complete results provided in the Appendix~\cite{gab}.

\begin{figure}[t]\centering

	\scalebox{0.2}[0.2]{\includegraphics{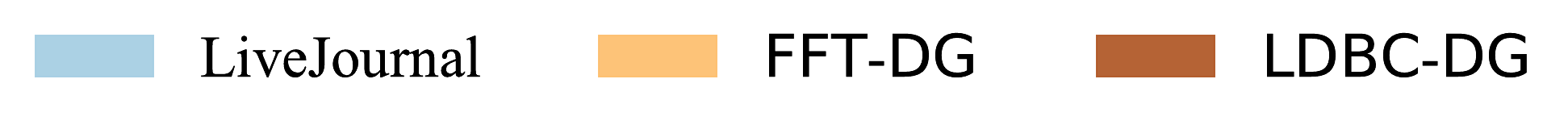}}

	\vspace{-0.5em}

	\begin{subfigure}[b]{0.36\textwidth}
        \includegraphics[width=\textwidth]{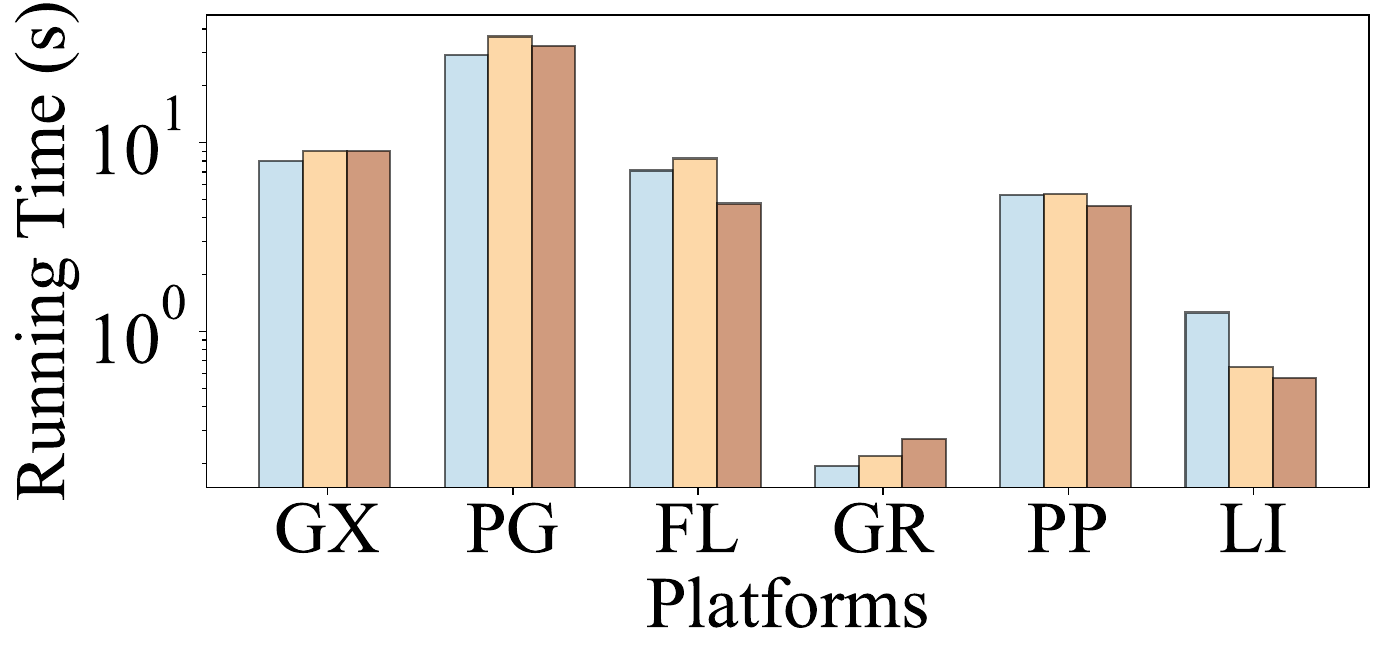}
    \end{subfigure}
	\begin{subfigure}[b]{0.36\textwidth}
        \includegraphics[width=\textwidth]{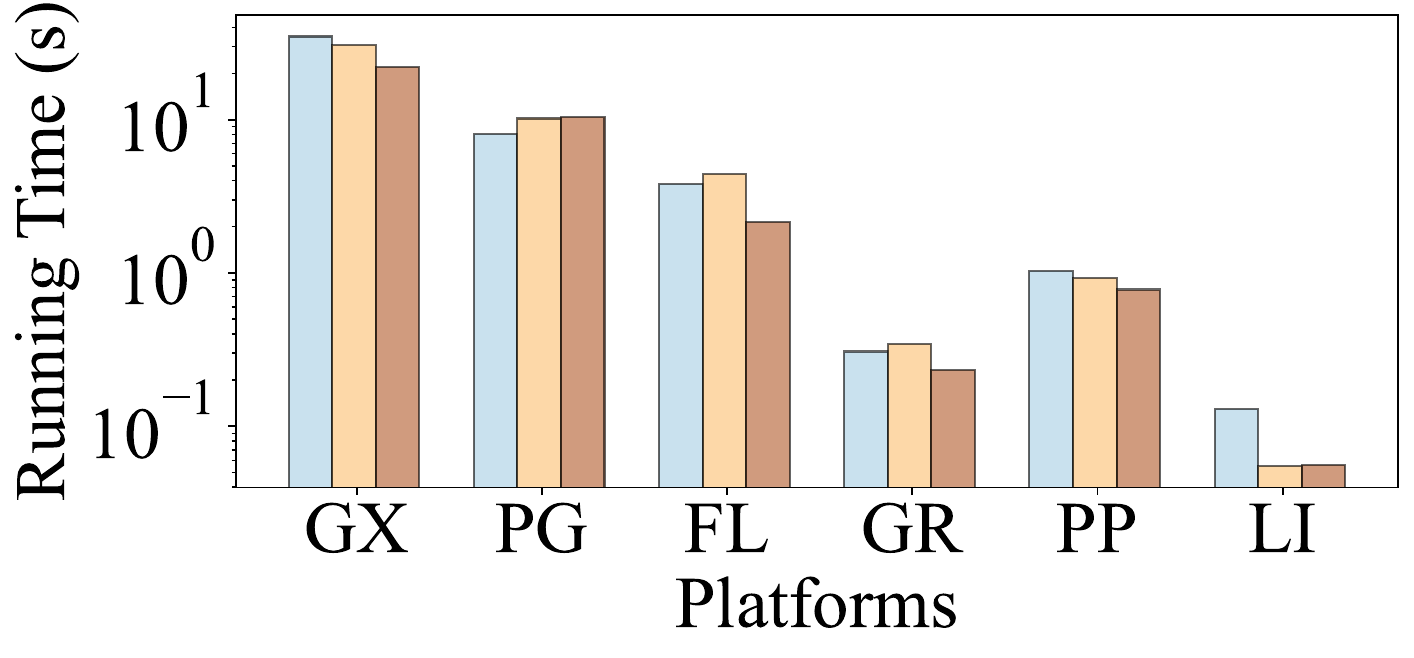}
    \end{subfigure}

	\vspace{-0.8em}

	\caption{Running time of \pr and \sssp on three datasets}
	\label{fig:exp_generator_time}
\end{figure}

\notation{D.(2)} \revise{We also generate \fftdg and \ldbcdg datasets with the same size as the LiveJournal and evaluate the performance of \pr and \sssp on all six supported platforms. For \fftdg, we tune the density factor $\alpha$, and for \ldbcdg, we reduce the degree of all vertices. The performance and relative difference are shown in Figure~\ref{fig:exp_generator_time} and Table~\ref{tab:exp_relative_diff}, respectively. 
Compared to \ldbcdg, the \fftdg dataset shows equal (or slightly better) runtime similarity to the real-world dataset (LiveJournal) on most platforms, with a difference within 25\% (except for Ligra).
In fact, the total workload on distributed platforms is quite unpredictable so this evaluation can validate the performance similarity.}

\begin{table}[t!]
	\centering
	\caption{Jensen-Shannon Divergence between LiveJournal and \fftdg / \ldbcdg 's datasets}
    \vspace{-0.8em}
	\label{tab:JS}
	\small
	\begin{tabular}{c|c|c|c|c|c|c}
		\hline
		\textbf{Generator} & \textbf{CC} & \textbf{TPR} & \textbf{BR} & \textbf{Diam} & \textbf{Cond} & \textbf{Size} \\ \hline
		\fftdg              & 0.108       & 0.057        & 0.057       & 0.053         & 0.054         & 0.069         \\ \hline
		\ldbcdg              & 0.201       & 0.097        & 0.087       & 0.236         & 0.183         & 0.133         \\ \hline
	\end{tabular}
 \vspace{-0.4cm}
\end{table}

\begin{figure}[t!]\centering
	\hspace{-3em}
	\begin{subfigure}[b]{0.24\textwidth}
        \includegraphics[width=\textwidth]{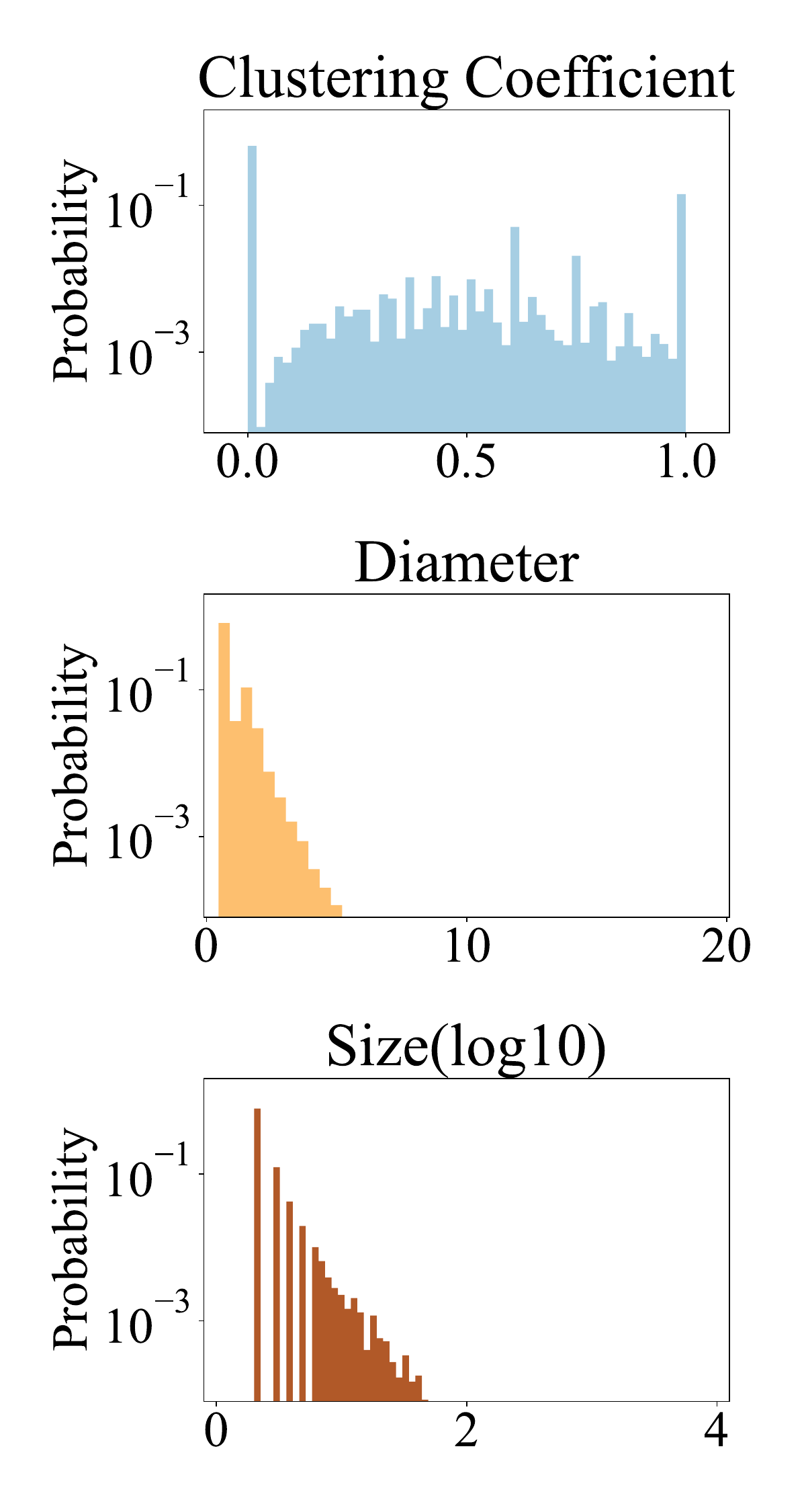}
		\vspace{-2em}
		\caption{LiveJournal}
    \end{subfigure}
	\hspace{-1em}
	\begin{subfigure}[b]{0.24\textwidth}
        \includegraphics[width=\textwidth]{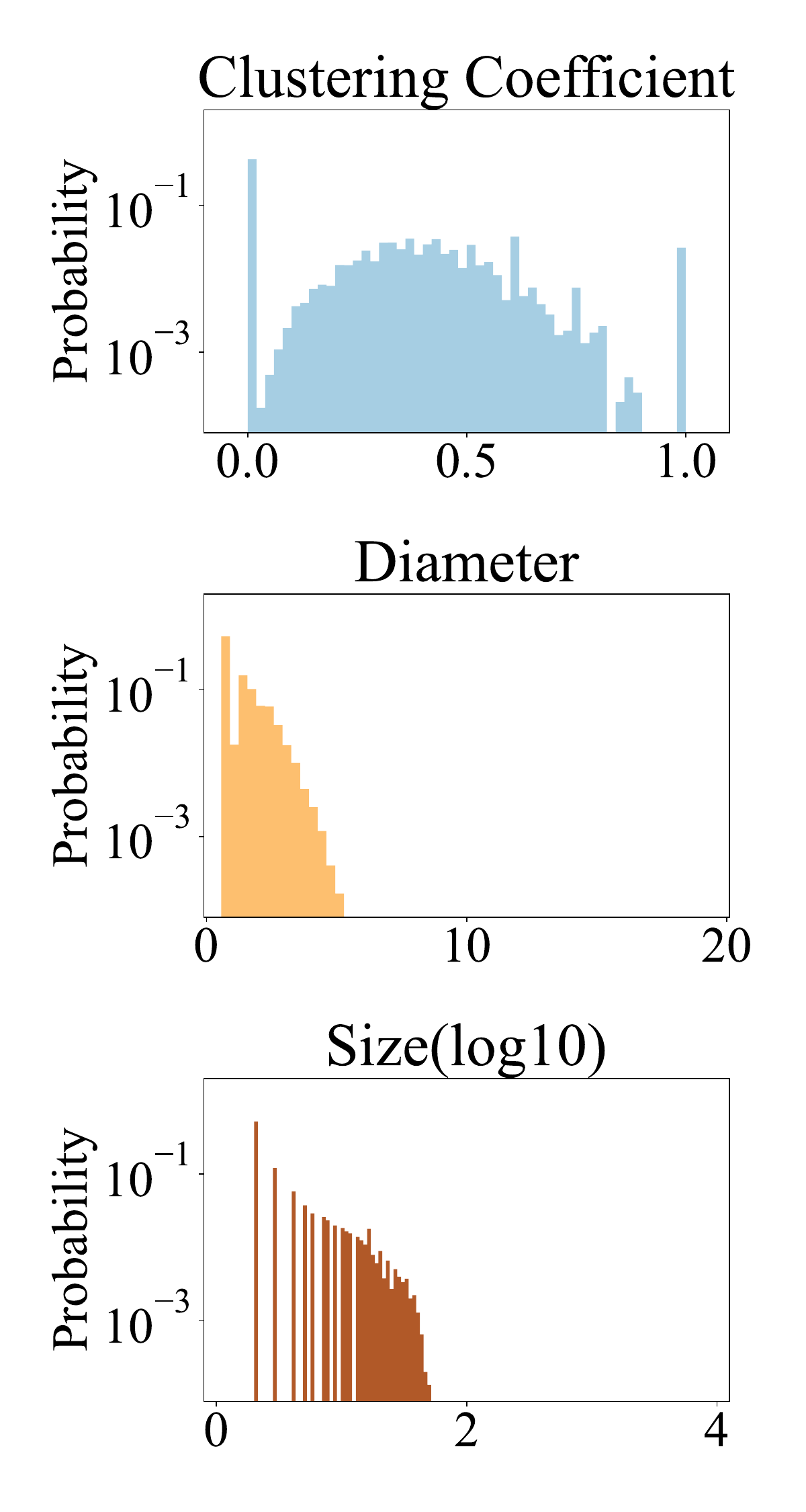}
		\vspace{-2em}
		\caption{\fftdg}
    \end{subfigure}
	\hspace{-1em}
	\begin{subfigure}[b]{0.24\textwidth}
        \includegraphics[width=\textwidth]{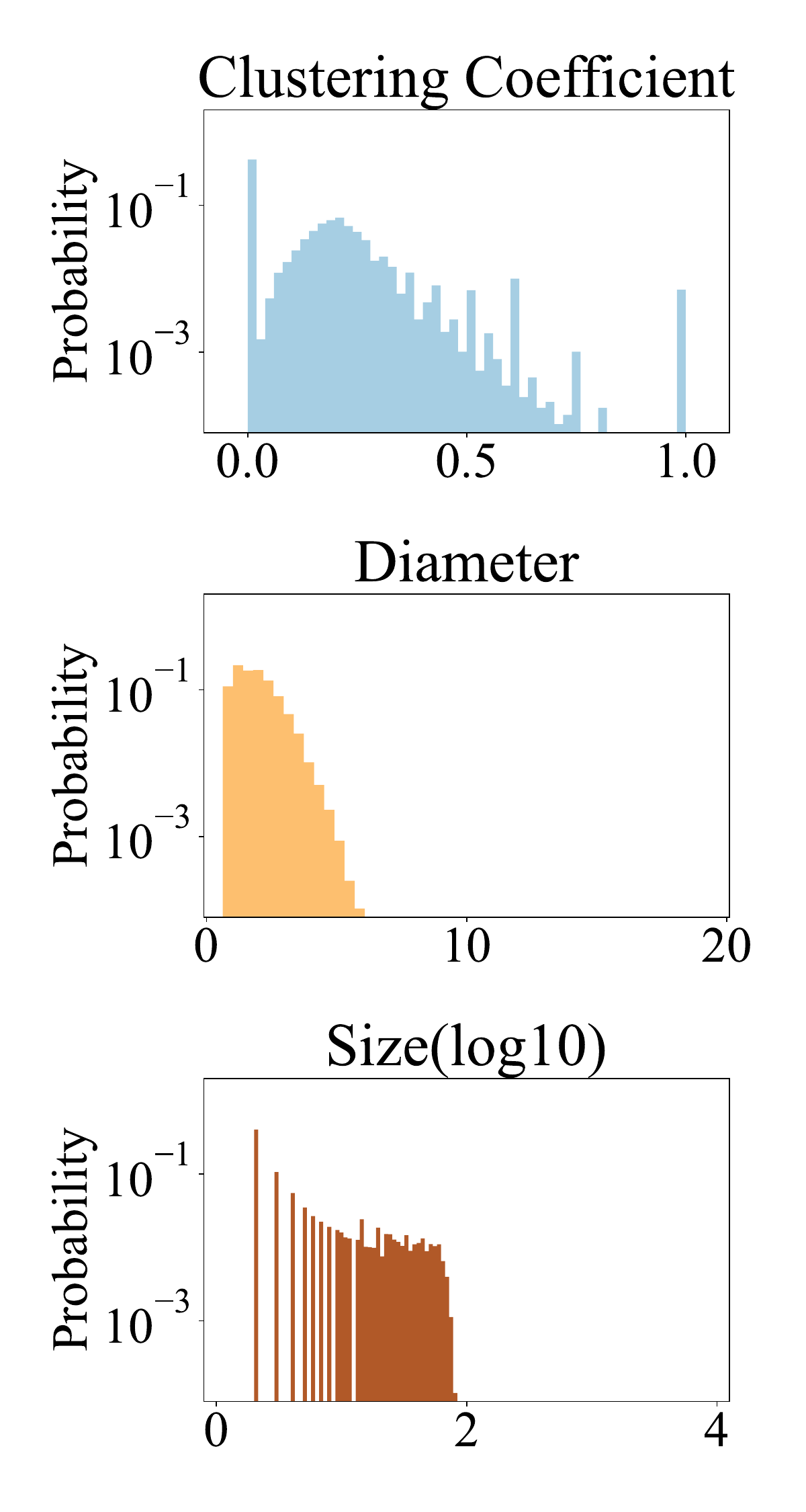}
		\vspace{-2em}
		\caption{\ldbcdg}
    \end{subfigure}
	\hspace{-3em}

	\vspace{-0.8em}

	\caption{Distribution of dataset communities statistics}
	\label{fig:generator}
\end{figure}

\begin{table}[t]
	\centering
	\caption{Relative Difference between LiveJournal and \fftdg / \ldbcdg 's datasets}\vspace{-0.8em}
	\label{tab:exp_relative_diff}
	\small
	\begin{tabular}{c|c|c|c|c|c|c|c}
		\hline
		\textbf{Algo.}         & \textbf{Generator} & \textbf{GX} & \textbf{PG} & \textbf{FL} & \textbf{GR} & \textbf{PP} & \textbf{LI} \\ \hline
		\multirow{2}{*}{\pr} & \fftdg             & 12\%             & 25\%                 & 15\%            & 12\%            & 1\%              & 48\%            \\
								   & \ldbcdg            & 12\%             & 11\%                 & 33\%            & 38\%            & 11\%             & 57\%            \\ \hline
		\multirow{2}{*}{\sssp}     & \fftdg             & 11\%             & 25\%                 & 15\%            & 11\%            & 10\%             & 57\%            \\
								   & \ldbcdg            & 37\%             & 28\%                 & 43\%            & 24\%            & 24\%             & 56\%             \\ \hline
		\end{tabular}
\end{table}

\begin{figure}[t!]
	\centering
	\scalebox{0.42}[0.42]{\includegraphics{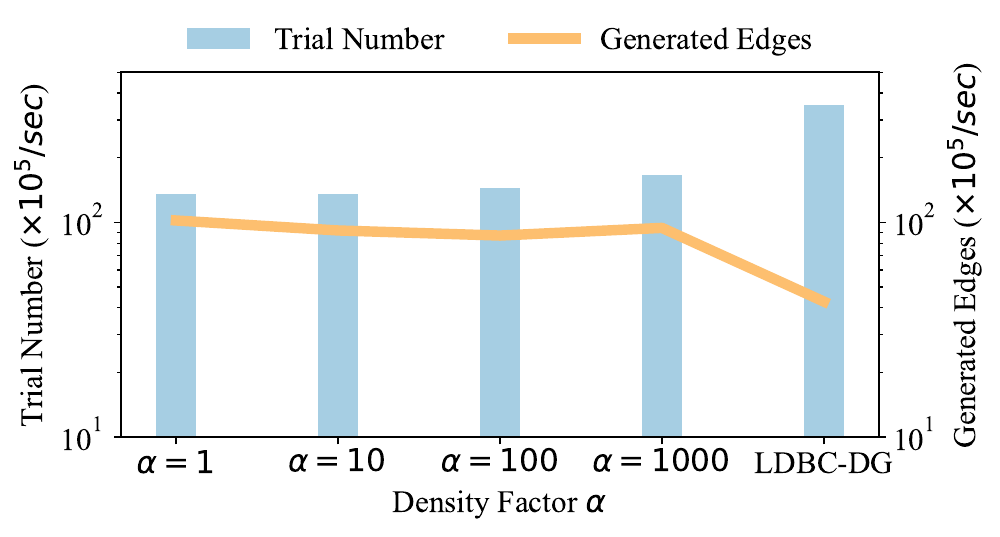}}
	\vspace{-1em}
	\caption{Generated edge number and generating time with different density factor $\alpha$.}
	\label{fig:exp_density}
	\vspace{-0.5em}
\end{figure}

\stitle{Generation Efficiency.} The efficiency evaluation involves generating multiple datasets with the density factor $\alpha$ varying in $\{1, 10, 100, 1000\}$. Figure~\ref{fig:exp_density} illustrates the total number of trials and generated edges per second. \fftdg demonstrates consistent throughput as density increases, requiring about $1.5$ trials per edge, while \ldbcdg necessitates more than $8$ trials per edge. Although \ldbcdg exhibits a faster sampling process, capable of executing up to 2 times the number of trials per second, its edge generation speed is still about 2.2 times slower than \fftdg.

\subsection{Algorithm \& Statistics Impact}
\label{sec:7.1}

\begin{figure*}[t]\centering


	\begin{subfigure}[b]{0.4\textwidth}
        \includegraphics[width=\textwidth]{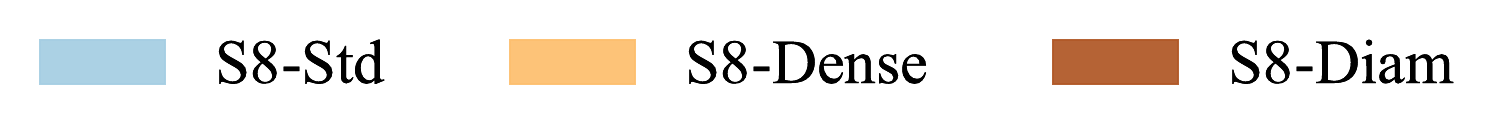}
    \end{subfigure}


	\begin{subfigure}[b]{0.48\textwidth}
		\includegraphics[width=\textwidth]{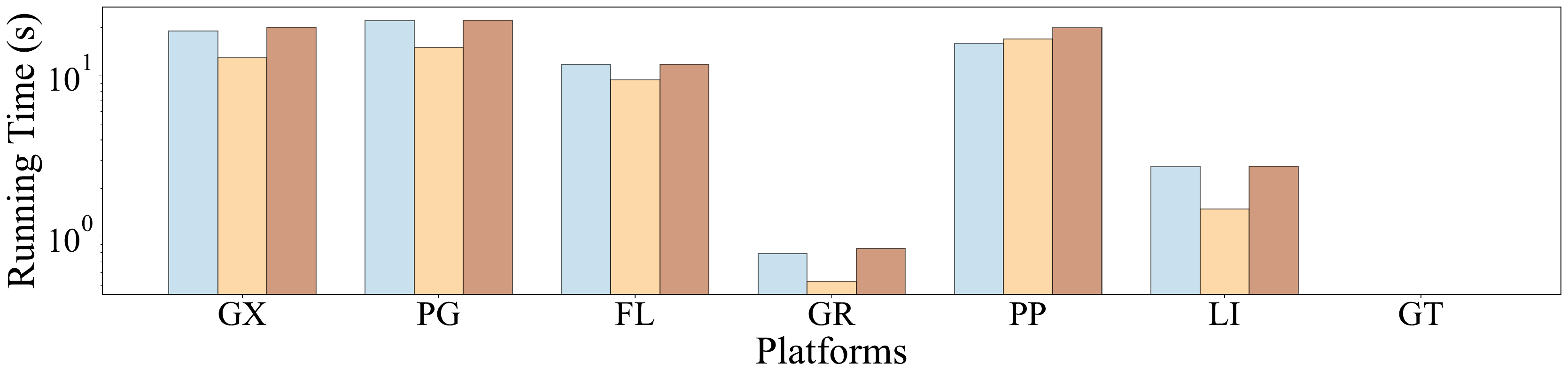}
		\vspace{-1.8em}\caption{\footnotesize \pr~(Iterative)}
	\end{subfigure}
	\hspace{0.5em}
	\begin{subfigure}[b]{0.48\textwidth}
		\includegraphics[width=\textwidth]{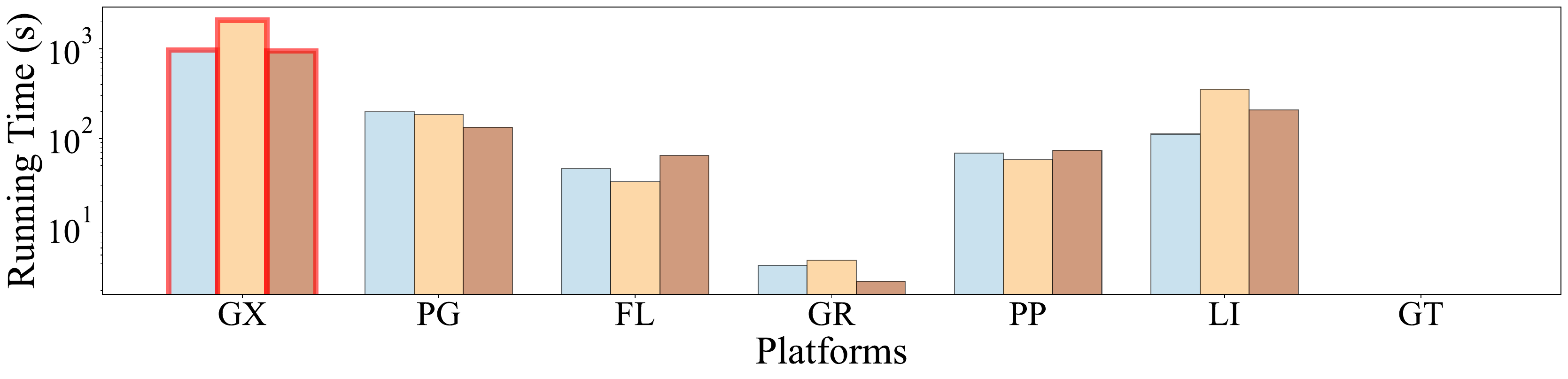}
		\vspace{-1.8em}\caption{\footnotesize{\lpa~(Iterative)}}
	\end{subfigure}

	\begin{subfigure}[b]{0.48\textwidth}
		\includegraphics[width=\textwidth]{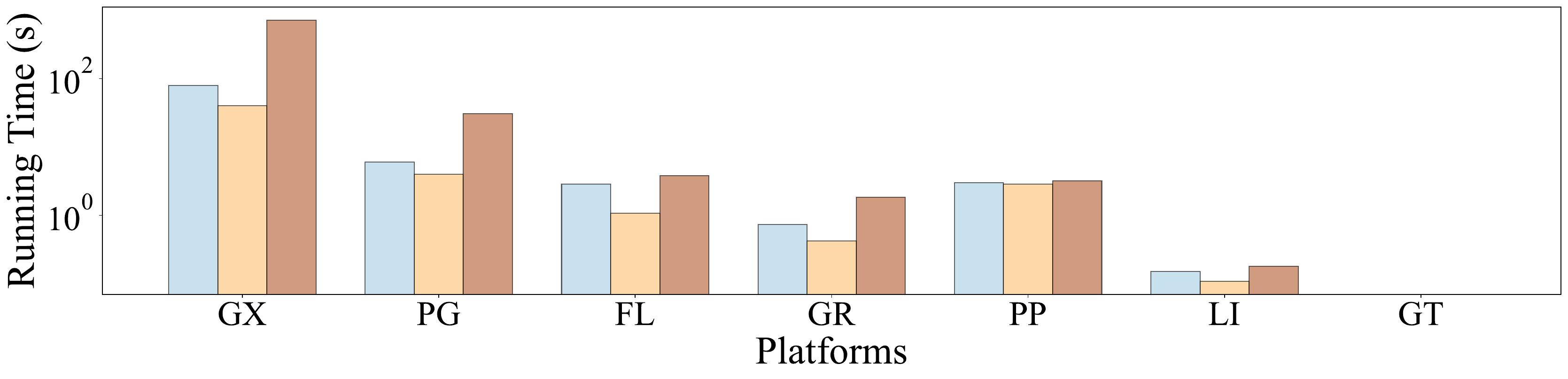}
		\vspace{-1.8em}\caption{\footnotesize{\sssp~(Sequential)}}
	\end{subfigure}
	\hspace{0.5em}
	\begin{subfigure}[b]{0.48\textwidth}
		\includegraphics[width=\textwidth]{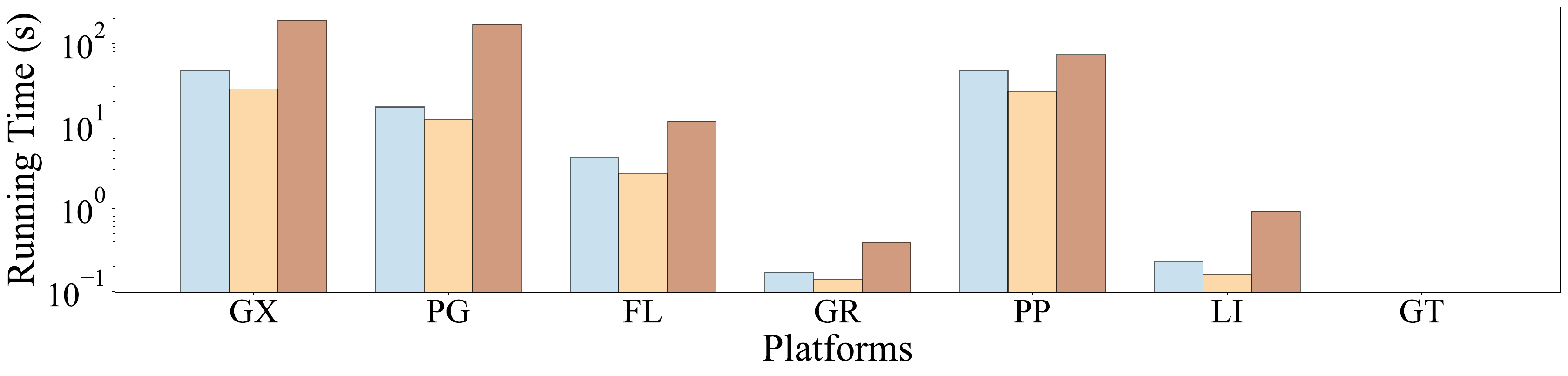}
		\vspace{-1.8em}\caption{\footnotesize{\cc~(Sequential)}}
	\end{subfigure}

	\begin{subfigure}[b]{0.48\textwidth}
		\includegraphics[width=\textwidth]{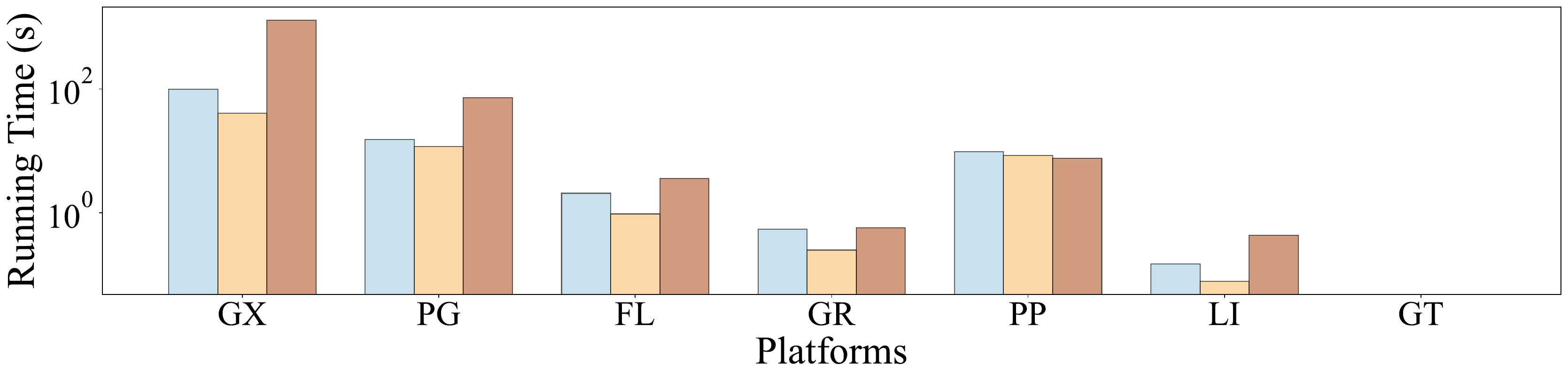}
		\vspace{-1.8em}\caption{\footnotesize{\bc~(Sequential)}}
	\end{subfigure}
	\hspace{0.5em}
	\begin{subfigure}[b]{0.48\textwidth}
		\includegraphics[width=\textwidth]{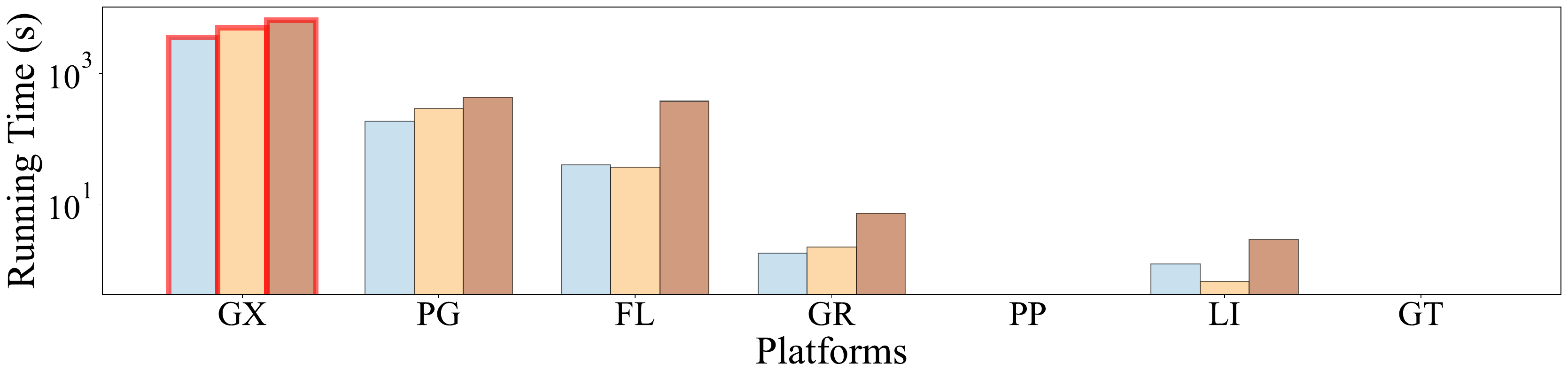}
		\vspace{-1.8em}\caption{\footnotesize{\core~(Sequential)}}
	\end{subfigure}

	\begin{subfigure}[b]{0.48\textwidth}
		\includegraphics[width=\textwidth]{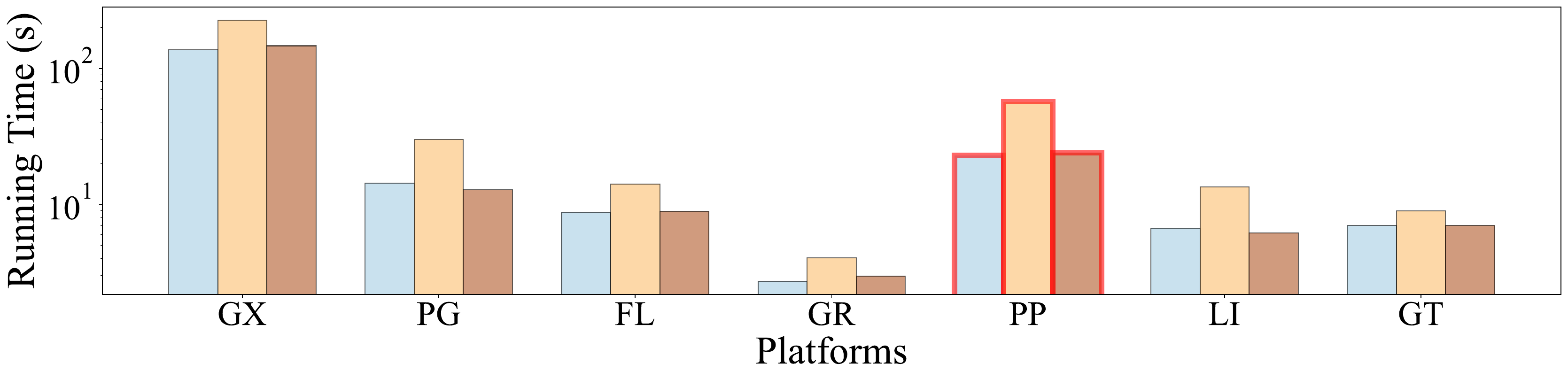}
		\vspace{-1.8em}\caption{\footnotesize{\tc~(Subgraph)}}
	\end{subfigure}
	\hspace{0.5em}
	\begin{subfigure}[b]{0.48\textwidth}
		\includegraphics[width=\textwidth]{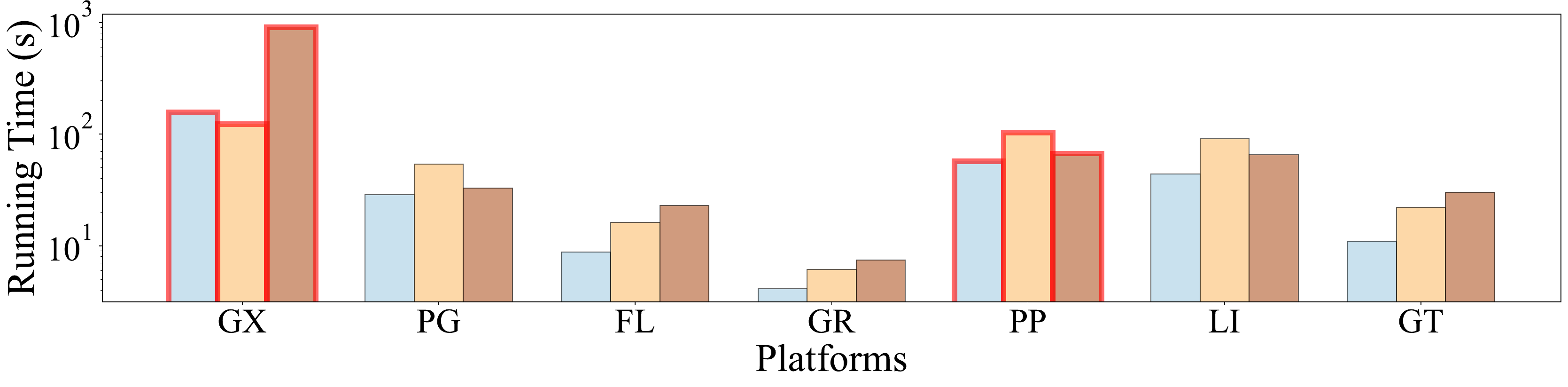}
		\vspace{-1.8em}\caption{\footnotesize{\clique~(Subgraph)}}
	\end{subfigure}

	\vspace{-1em}
	\caption{Running time of eight algorithms on datasets S8-Std, S8-Dense, S8-Diam}
	\label{fig:exp_algorithm_sensitivity}
	\vspace{-1.5em}
\end{figure*}

We present experimental results of algorithms and statistics impact in Figure~\ref{fig:exp_algorithm_sensitivity} and analyze them collectively.

\notation{D.(5)} \revise{\stitle{Algorithm Coverage across Platforms.} All experiment codes are sourced from official documentation, third-party repositories, or implemented by us to the best of our ability. Out of the total $8$ (algorithms) $\times$ $7$ (platforms) $=56$ cases, we successfully evaluate 49 cases.
An algorithm’s suitability for a given computing model (Section \ref{sec:algorithm_suitability}) affects whether it is feasible and/or efficient to implement the algorithm on different platforms.
Among the 7 unimplemented cases: \pregel's interface lacks support for managing variables like coreness values required by \cd across supersteps; \pr, \lpa, \sssp, \wcc, \bc, and \cd cannot be implemented on \gthinker, a subgraph-centric platform that does not support the iterative control flow needed by these algorithms.
Additionally, \graphx struggles with \lpa, \cd, and \kc due to the overhead for adapting the Spark RDD APIs; \pregel lacks push/pull optimizations found in newer platforms like \flash. Consequently, it faces high computation overhead with subgraph algorithms (\tc and \kc). To manage these inefficiencies, we use 16 machines instead of one single machine for these cases (indicated by \underline{\textbf{red bars}} in Figure~\ref{fig:exp_algorithm_sensitivity}).}


Generally, eight algorithms exhibit three computation patterns:


\stitle{Iterative Algorithms~(\pr and \lpa)} perform similarly on S8-Std and S8-Diam, suggesting that these algorithms are not sensitive to diameter. However, there are noticeable differences in performance on S8-Dense.
For \pr, all platforms run faster on S8-Dense as expected, as \pr distributes its workload across vertices, and the denser dataset has fewer vertices with the same number of edges.
The performance of \lpa is affected by implementation, which requires hash tables to store vertex labels. Specifically, \graphx experiences significant performance degradation when merging hash tables from different vertices. In contrast, other platforms maintain a local hash table and can merge received hash tables into the local one, avoiding much redundant computation.

\stitle{Sequential Algorithms~(\sssp, \wcc, \bc, \cd)} generally perform better on S8-Dense but worse on S8-Diam, indicating sensitivity to both density and diameter. Similar to \pr, workloads are distributed over vertices, resulting in improved performance on S8-Dense. However, their sequential nature and the increased synchronization requirements in S8-Diam result in reduced performance.

For \sssp and \bc, both \vc and \ec platforms like \graphx and \power cost too much time on multiple synchronizations, while the \blc platform like \grape is not sensitive to the diameter. Since only a portion of vertices are active in each iteration, \vc optimizations like the push-pull model and the vertex subset technique in \flash and \ligra can also improve efficiency. Besides, \pregel performs equally well on three datasets, showing a strong workload balance.

\notation{D.(4)} \revise{\flash and \pregel offer standard Pregel-like APIs that support global communication, i.e., send messages to any vertices, enabling some advanced algorithms such as HashMin~\cite{rastogi2013finding} and Shiloach-Vishkin~\cite{vishkin1982log, yan2014pregel, qin2014scalable} for \wcc and reducing iteration rounds significantly.} \blcc models like \grape can even directly call sequential implementation, e.g., Disjoint Set, for dynamic \wcc. In contrast, some \vc platforms (e.g., \graphx) and the \ec model can only send messages to neighbors.

Unlike other sequential algorithms, the performance of \core exhibits varying trends across different platforms when executed on denser datasets. Since the coreness value could be very large (especially in S8-Dense) and multiple iterations may be required for each coreness value, \graphx is extremely slow even when using 16 machines. \revise{Additionally, for each coreness value, platforms like \graphx and \power need to activate all vertices, while \flash and \ligra can maintain the activated vertices, resulting in better performance.}

\stitle{Subgraph Algorithms~(\tc, \kc)} require substantial communication to broadcast neighbor lists and massive computation to count subgraphs, where \blc (\grape) and \sgc (\gthinker) models can perfectly reduce overhead. For \tc, all platforms count the common neighbors of two vertices of each edge, where denser datasets have larger neighbor lists, resulting in poorer performance on S8-Dense. Unlike \tc, the performance of \clique on S8-Diam is also poor. The reason is that to generate a long-diameter graph with the same scale, the local partition of the graph must be relatively dense, leading to substantially more cliques. Therefore, the bottleneck is communication for \power, \pregel, \ligra, but computation for \graphx, \flash, \grape, \gthinker.

\subsection{Scalability Sensitivity}
\label{sec:7.2}

\noindent\textbf{Varying Number of Threads.} We first examine the scale-up performance on a single machine with increasing threads: $1$, $2$, $4$, $8$, $16$, and $32$. Since \graphx requires at least four threads for \pr and two threads for \sssp to operate effectively, we exclude only \tc.

\begin{figure}[t]\centering

	\scalebox{0.23}[0.23]{\includegraphics{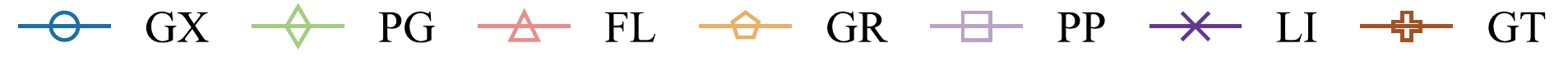}}

	\begin{subfigure}[b]{0.6\textwidth}
        \includegraphics[width=\textwidth]{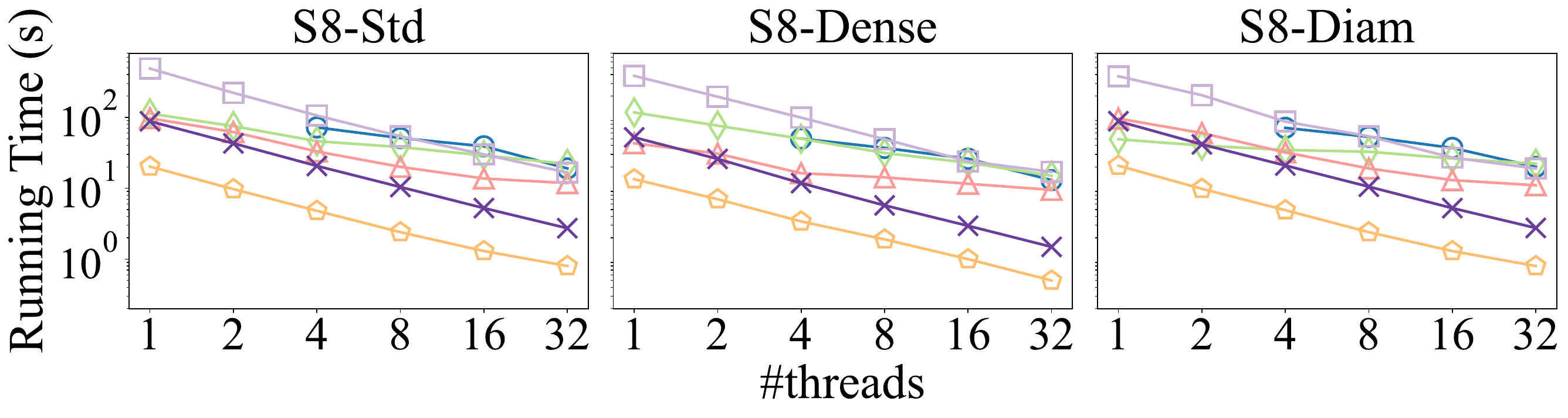}
    \end{subfigure}


	\begin{subfigure}[b]{0.6\textwidth}
        \includegraphics[width=\textwidth]{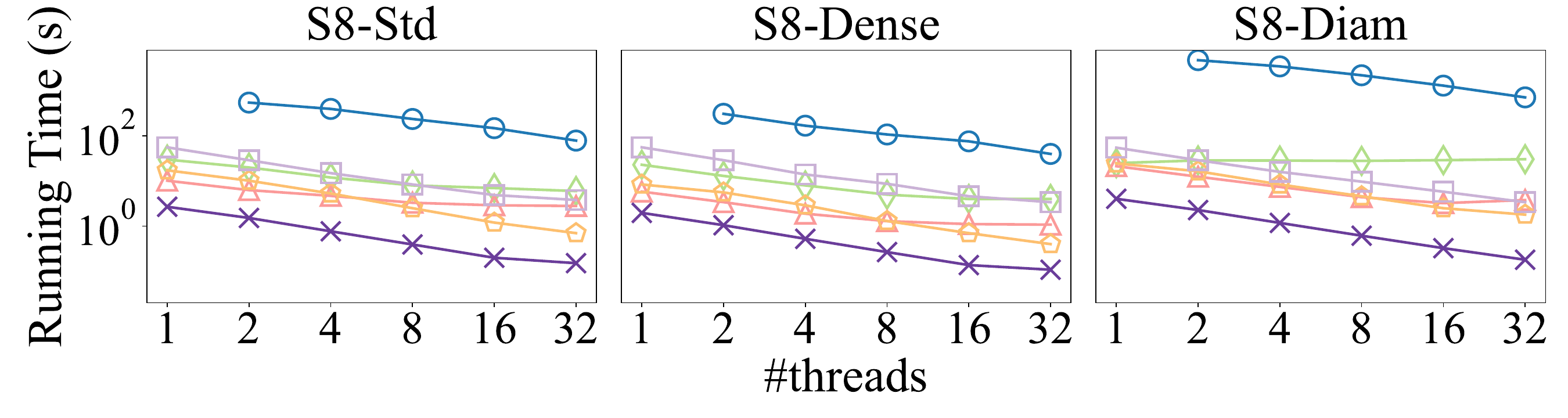}
    \end{subfigure}


	\begin{subfigure}[b]{0.6\textwidth}
        \includegraphics[width=\textwidth]{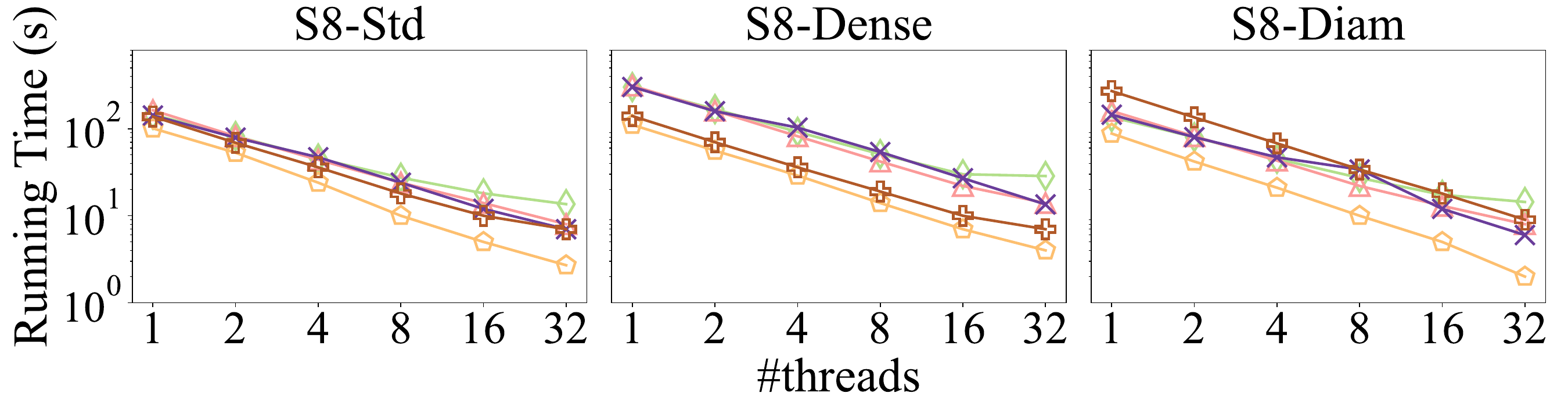}
    \end{subfigure}

	\vspace{-1em}
	\caption{Running time varying \#threads (Scale = 8)}
	
	\label{fig:exp_scalability_threads}
	\vspace{-0.5cm}
\end{figure}

Figure~\ref{fig:exp_scalability_threads} illustrates the scale-up performance of \pr, \sssp, and \tc across platforms. All platforms exhibit a clear decreasing trend in execution time as more threads are used. The detailed scaling factors, calculated as the ratio of best performance to single-thread performance, are presented in Table~\ref{tab:exp_scalability_threads}. \grape achieves the best scale-up performance, with a scaling factor of up to $37\times$, followed by \ligra, \gthinker and \pregel, which have an average scaling factor of $24.9\times$, $23.3\times$ and $19.6\times$ . The scaling factors for \flash, \graphx and \power range between $1\times$ and $12\times$. 

\begin{table}[t!]
    \caption{Speed up (threads)}
    \label{tab:exp_scalability_threads}
    \vspace{-1em}
	\small
    \begin{tabular}{c|c|c|c|c|c|c|c|c}
    \hline
        \textbf{\hspace{-0.2em}Algo.\hspace{-0.2em}}           &  \textbf{Dataset}   & \textbf{GX}  &  \textbf{PG}  &  \textbf{FL} &  \textbf{GR}  &  \textbf{PP}  &  \textbf{LI} &  \textbf{GT} \\ \hline
    \multirow{3}{*}{PR}   & S8-Std      & 3.8    & 5.1    & 8.2   & 25.3  & 29.3    & 32.2 & ---  \\
                        & S8-Dense    & 3.8    & 7.8    & 4.5   & 25.2  & 22.6    &  34.9 & ---  \\
                        & S8-Diam     & 3.6    & 2.2    & 8.2   & 24.2  & 18.9    & 32.0  & ---  \\ \hline
    \multirow{3}{*}{SSSP} & S8-Std      & 6.9    & 5.0    & 9.3   & 23.5  & 14.7     & 17.8 & ---   \\
                        & S8-Dense    & 7.8    & 5.8    & 8.5   & 19.7  & 16.4    & 18.2  & ---  \\
                        & S8-Diam     & 6.7    & 0.9    & 10.2  & 13.2  & 15.8    & 22.4 & ---   \\ \hline
    \multirow{3}{*}{TC}   & S8-Std      & ---    & 10.7   & 18.7  & 37.2  & ---    & 21.3 &19.7  \\
                        & S8-Dense    & ---    & 10.5   & 22.2  & 27.5  & ---    & 22.4 & 30.1 \\
                        & S8-Diam     & ---    & 9.4    & 18.1  & 29.6  & ---    & 23.5 & 20.0\\ \hline
    \end{tabular}
\end{table}

The scale-up performance varies significantly across the three algorithms. The scaling factor for \tc is higher than \pr and \sssp. This confirms our previous categorization: \tc requires no synchronization, allowing all computations to be parallelized, while \pr consists of $k$ iterations and $k$ synchronizations. Sequential algorithms like \sssp require multiple synchronizations, making their scalability the worst among the three algorithms. 
Besides, the scale-up performance is sensitive to dataset characteristics. Most platforms scale better on S8-Dense but perform worse on S8-Diam than on S8-Std. 

\begin{figure}[t]\centering


	\scalebox{0.23}[0.23]{\includegraphics{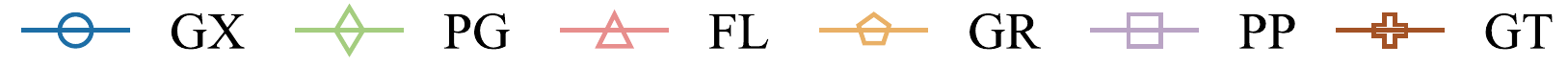}}

	\begin{subfigure}[b]{0.6\textwidth}
        \includegraphics[width=\textwidth]{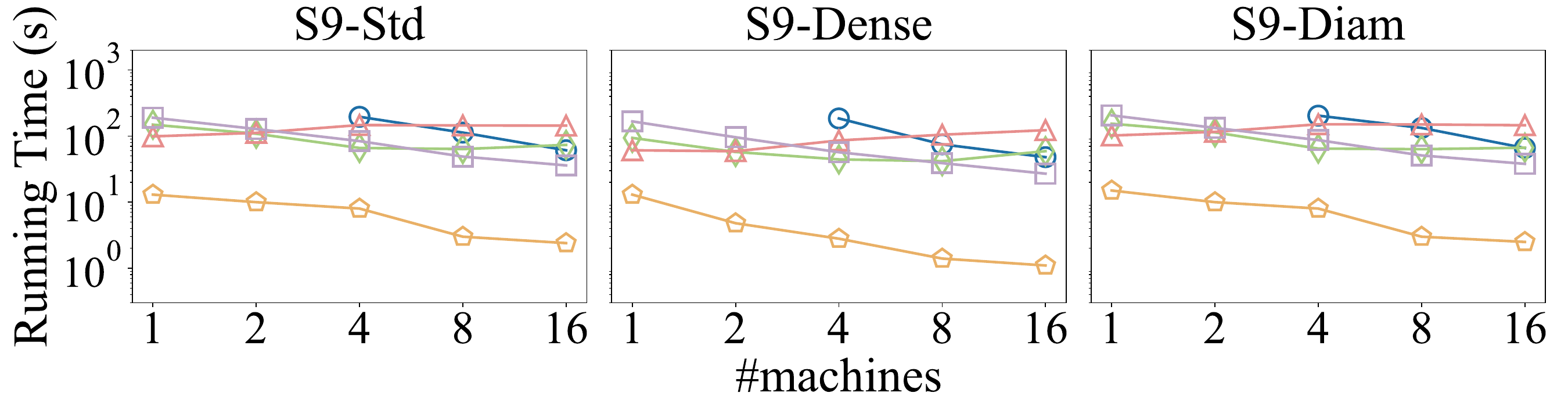}
    \end{subfigure}


	\begin{subfigure}[b]{0.6\textwidth}
        \includegraphics[width=\textwidth]{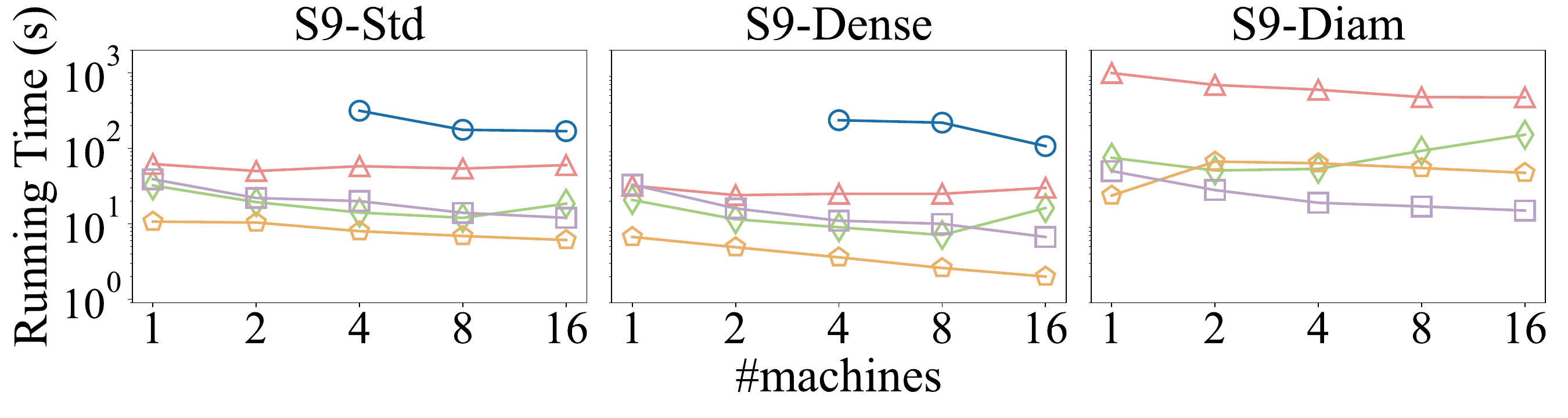}
    \end{subfigure}


	\begin{subfigure}[b]{0.6\textwidth}
        \includegraphics[width=\textwidth]{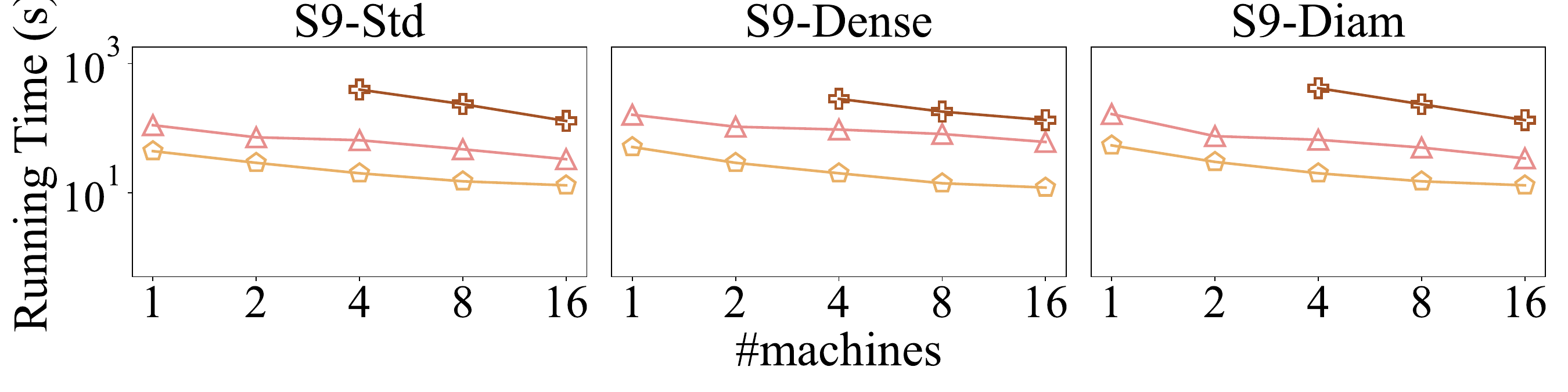}
    \end{subfigure}

	\vspace{-1em}

	\caption{Running time varying \#machines (Scale = 9)}
	\label{fig:exp_scalability_machines_9}
	\vspace{-0.5em}
\end{figure}

\noindent\textbf{Varying Number of Machines.} We evaluate scale-out performance by increasing the number of machines to $2$, $4$, $8$, and $16$, using larger S9 datasets. \graphx fails to compute \sssp on S9-Diam, and \graphx and \power encounter Out-Of-Memory errors on \tc. 
\ligra is not tested as it runs only on a single machine.
Figure~\ref{fig:exp_scalability_machines_9} shows the performance, and scaling factors, calculated as the ratio of best to single-machine performance, are summarized in Table~\ref{tab:exp_scalability_machines}.

\begin{table}[t]
	\vspace{0.2em}
   \caption{Speed up (machines)}
   \label{tab:exp_scalability_machines}
   \vspace{-1em}
   \small
	   \begin{tabular}{c|c|c|c|c|c|c|c}
		   \hline
		   \textbf{Algo.} &  \textbf{Dataset}   & \textbf{ GX } & \textbf{ PG }  & \textbf{ FL }  &  \textbf{ GR }  &  \textbf{ PP }  &  \textbf{ GT }  \\ \hline
		   \multirow{3}{*}{PR}   & S9-Std      & 3.2    & 2.3    & 0.8   & 5.8   & 5.2 & ---  \\
		   & S9-Dense    & 3.8    & 2.2    & 1.0   & 11.5  & 6.1 & ---  \\
		   & S9-Diam     & 3.0    & 2.4    & 0.8   & 6.1   & 5.4 & ---  \\ \hline
		   \multirow{3}{*}{SSSP} & S9-Std      & 1.8    & 2.6    & 1.2   & 1.7   & 3.2 & ---  \\
		   & S9-Dense    & 2.2    & 2.9    & 1.3   & 3.3   & 5.0  & --- \\
		   & S9-Diam     & ---    & 1.4    & 2.0   & 0.5   & 3.9  & --- \\ \hline
		   \multirow{3}{*}{TC}   & S9-Std      & ---    & ---    & 3.3   & 3.2   & --- & 3.1 \\
		   & S9-Dense    & ---    & ---    & 2.6   & 4.1   & ---  & 2.7 \\
		   & S9-Diam     & ---    & ---    & 4.7   & 3.9   & --- & 3.1\\ \hline
	   \end{tabular}
   \vspace{-0.5em}
\end{table}

All platforms show worse scale-out performance than scale-up, likely due to the high overhead of inter-machine communication. \pregel performs significantly better in scale-out than the others. Notably, \grape performs well on a single machine, but its performance gains saturation when scaling to multiple machines due to communication overhead. The scalability trends, influenced by algorithms and dataset characteristics, are similar to those in the scale-up experiments and won't be further discussed.




\subsection{Usability Evaluation}
\label{sec:7.5}

\begin{figure}[t]
	\begin{center}
		\begin{tabular}[t]{c}
			\includegraphics[width=0.65\columnwidth]{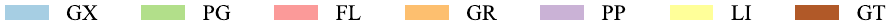}
		\end{tabular}

		\vspace{-0.3em}

		\begin{tabular}[t]{c}
			\includegraphics[width=0.75\columnwidth]{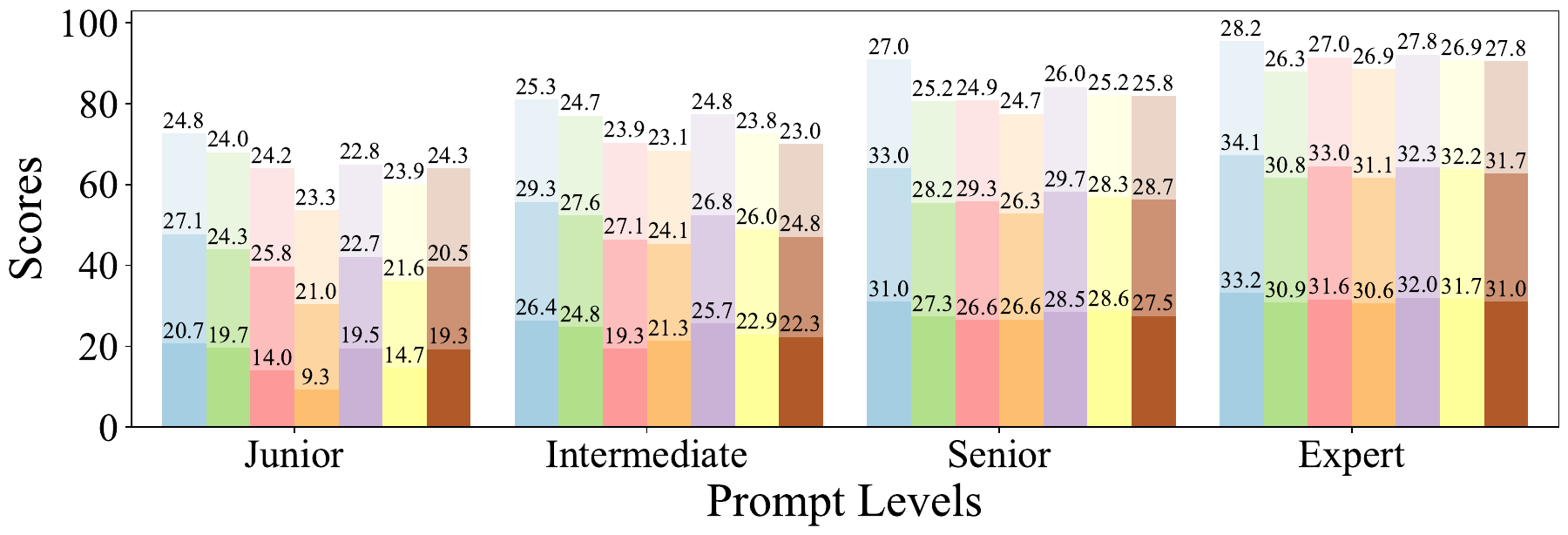}
		\end{tabular}
	\end{center}
	\vspace{-1.2em}
	\caption{Usability scores of varying platforms. The colors from deeper to lighter shades represent the scores for Compliance, Correctness, and Readability, respectively.}
	\label{usability_scores}
	\vspace{-0.5em}
\end{figure}

We use our usability evaluation framework to simulate programmers with varying expertise levels and evaluate the usability of APIs by analyzing the generated code.
Figure~\ref{usability_scores} shows the usability scores for various platforms with different prompt levels (from junior to expert). \revise{To validate our framework’s effectiveness, we invited over 80 students and developers with varying levels of familiarity with graph algorithms to assess the quality of code generated by our \textit{Code Generator} at the ``Intermediate'' and ``Senior'' prompt levels. Platform-related information was masked for the human reviewers to ensure unbiased evaluations. 
Table~\ref{expert_evaluation} compares our framework's results with the human reviews.}

From Figure~\ref{usability_scores}, we see that \graphx achieves the highest scores across all expertise levels, with particularly outstanding scores from senior and expert users, indicating that its API design is highly user-friendly, providing an excellent experience for both novice and advanced users. \power and \pregel show balanced and relatively high usability scores, especially among junior and intermediate users, suggesting their APIs are intuitive for beginners. 
In contrast, \grape has the lowest scores from junior users, reflecting a steep learning curve and a less beginner-friendly design. However, its scores rise significantly for senior and expert users, indicating it offers robust functionality once users gain experience.
Finally, \flash, \ligra, and \gthinker share similar patterns—lower scores at the junior level and higher scores at the senior and expert levels.
While their abstraction of graph traversal interfaces can simplify programming, \mbox{it may cause comprehension difficulties for beginners.}


\revise{The results from human reviewers (Table~\ref{expert_evaluation}) show a similar relative ranking to those of our framework. \graphx consistently receives the highest ranking, while \grape is ranked the lowest, with other platforms ranking in between.
}

\begin{table}[t!]
	\centering
	 \small
	 \caption{\revise{Comparison of code evaluation results: LLM vs. human evaluations from over 80 students and developers (numbers in parentheses represent rankings)}}
     \vspace{-0.8em}
		\begin{tabular}{c|c|c|c|c|c|c|c|c}
			\hline
			\textbf{Prompt}  & \textbf{Eval.} & \textbf{GX}  &  \textbf{PG}  & \hspace{-0.6em} \textbf{FL} \hspace{-0.6em} &  \textbf{GR}  &  \textbf{PP}  &  \textbf{LI}   &  \textbf{GT}  \\ \hline 
			\multirow{2}{*}{Intermediate} &	LLM       & 81.0(1) & 77.0(2)  & 70.3(5) & 68.5(7) & 73.3(3) & 72.7(4) & 70.0(6) \\ 
			                    &   Human   & 77.4(1)   & 62.8(5)   & 68.8(3)   & 57.2(7)   & 70.3(2)   & 67.6(4)  & 61.7(6) \\ \hline
			\multirow{2}{*}{Senior} &	LLM       & 91.0(1) & 80.6(6) & 80.8(5) & 77.5(7) & 84.2(2) & 82.1(3) & 82.0(4) \\ 
			                    &   Human   & 78.2(1)   & 61.6(5)   & 74.6(2)   & 56.8(7)   & 72.0(3)   & 72.0(4)  & 65.7(6)  \\ \hline
			
		\end{tabular}
	\label{expert_evaluation}
	\vspace{-0.5em}
\end{table}

\revise{ \stitle{\underline{Discussions of Scoring Difference.}}
Human evaluation relies on a five-point scale (1–5) to keep the process manageable for human reviewers. In contrast, our LLM-based framework applies more detailed evaluation rules with multiple dimensions, leading to finer-grained evaluation. As a result, the absolute scores produced by the LLM-based framework may differ from human evaluations. However, our focus is on the relative rankings and trends between LLM and human evaluation, rather than absolute score values. To assess the similarity of rankings between the two methods, we use Spearman’s Rho, a statistical measure that quantifies the degree of correlation between two sets of ranked data, which ranges from -1 to 1, with values closer to 1 indicating stronger consistency between the rankings.
The results, 0.75 for Intermediate and 0.714 for Senior, indicate that our framework’s ranking closely mirrors human judgment.
We observe cases where human rankings diverge from the LLM’s, such as in the evaluation of \power~ at the “Intermediate” level. We speculated that the LLM may favor \power~’s code due to its well-formulated APIs in this case. This presents an interesting future work to further fine-tune the LLM evaluator. }

\subsection{Summary}
\notation{D.(1)}\revise{The experimental evaluation verifies the effectiveness of our benchmark, indicating the performance of a platform is affected by the different algorithms and datasets. 
For instance, \grape and \gthinker perform well in handling subgraph algorithms, while \grape and \ligra excel in sequential algorithms. 
\pregel shows balanced performance across different datasets, while \graphx, \flash, and \power are more sensitive. Additionally, our usability evaluation examines platform API usability, revealing their friendliness to developers in daily use. Based on the trade-off between performance and usability, we offer valuable insights for users in platform selection, which will be presented in Section~\ref{sec:guide}.}

\section{Platform Selection Guide}
\label{sec:guide}

\begin{figure}[t!]\centering
	
	\begin{subfigure}[b]{0.6\textwidth}
		\includegraphics[width=\textwidth]{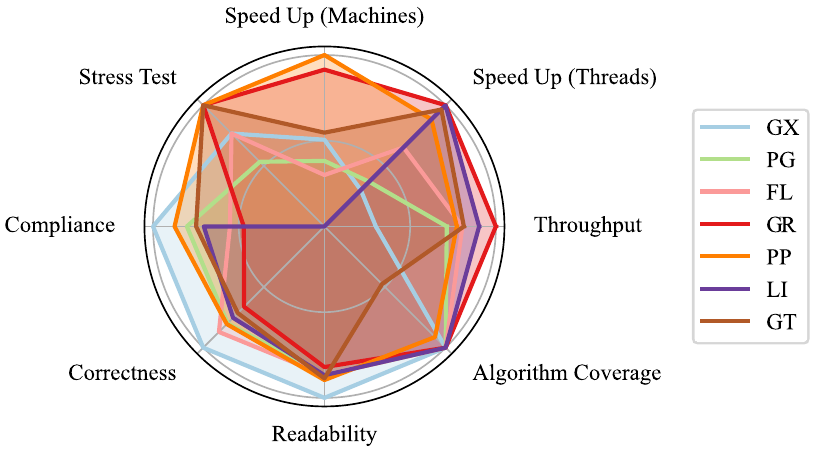}
	\end{subfigure}
	\vspace{-1em}
	\caption{Comprehensive comparison across platforms}
	\label{fig:exp_spider}
	\vspace{-0.8em}
\end{figure}

\revise{Based on the results in Section~\ref{sec:exp_result}, Figure~\ref{fig:exp_spider} shows the performance of all platforms across different metrics. According to the area covered by each platform, the ranking is as follows: \pregel $>$ \grape $>$ \graphx $>$ \gthinker $>$ \flash $>$ \power $>$ \ligra. 
\pregel stands out as the top-performing platform overall, performing well on all metrics without any notable weaknesses and ranking in the top three for nearly all categories.
\grape excels in throughput, machine speed-up, thread speed-up, and stress tests, showing strong performance and scalability. However, it scores the lowest in the three usability evaluation metrics, indicating a steep learning curve and complex API. In contrast, \graphx achieves the highest scores in compliance, correctness, and readability, highlighting its exceptional API usability. However, it performs poorly in thread speed-up, throughput, and stress test, indicating limited performance and scalability. 
\gthinker delivers a balanced performance in both usability and efficiency but supports fewer algorithms, leading to a lower algorithm coverage score and a mid-tier ranking. \flash and \power also show balanced performance, but their individual rankings for each metric are not particularly outstanding, resulting in lower overall rankings. 
\ligra performs well for single-machine tasks, especially in thread speed-up, but ranks lowest overall due to its lack of support for distributed computing.}

\notation{D.(7)}\revise{
	\graphx is ideal for users of all levels, as long as performance and scalability aren not top priorities, thanks to its excellent usability. \pregel, \flash and \power are also suitable for most users, especially beginners and intermediates, due to their balanced usability and good performance with large-scale data. 
For users requiring high performance, \gthinker and \ligra are strong choices. However, \gthinker supports fewer algorithms and is best suited for compute-intensive tasks, while \ligra provides robust interfaces and extensive algorithm support, excelling in single-machine tasks. Due to its lack of support for distributed computing, \ligra is unsuitable for large-scale or multi-machine workloads. 
For maximum performance and scalability, \grape is recommended despite its steeper learning curve, as it delivers significant efficiency once mastered.}

\section{Conclusion}
\label{sec:conclusion}
In this paper, we present a new benchmark for large-scale graph analytics platforms, featuring eight algorithms and the \fftdg, which improves dataset generation by adjusting scale, density, and diameter. We also introduce a multi-level LLM-based framework for evaluating API usability, marking the first use of such metrics in graph analytics benchmarks. Our extensive experiments assess both performance and usability, offering valuable insights for developers, researchers, and practitioners in platform selection.

\begin{acks}
We thank the 80 anonymous human experts for their valuable contributions to the usability evaluation.
Long Yuan is supported by National Key R\&D Program of China 2022YFF0712100, NSFC62472225.
Peng Cheng is supported by NSFC 62102149.
Xuemin Lin is supported by NSFC U2241211 and U20B2046.

\end{acks}

\bibliographystyle{ACM-Reference-Format}
\bibliography{main}


\end{document}